\documentclass[twocolumn,aps,floats,showpacs,amsmath]{revtex4}
\usepackage{amssymb,amsmath}
\usepackage{graphicx,float}

\begin{document}
\title{ \Large \bf  Spectral and Dynamical Properties in Classes of Sparse Networks with Mesoscopic Inhomogeneities}
\author{\large  \flushleft 
Marija Mitrovi\'c$^{\star,\diamond}$ and Bosiljka Tadi\'c$^{\diamond}$
} 

\affiliation{ %\flushleft
$^\star$ Scientific Computing Laboratory; Institute of Physics; 11000 Belgrade; Serbia\\
$^\diamond$Department  for Theoretical Physics; Jo\v{z}ef Stefan Institute; 
P.O. Box 3000; SI-1001 Ljubljana; Slovenia, 
\\ \hspace{1cm}} 

%\date{}
\begin{abstract}
We study structure, eigenvalue spectra and random walk dynamics in a wide class of networks with subgraphs (modules) at mesoscopic scale.
The networks are grown within the model with three parameters controlling the
number of modules, their internal structure as scale-free and
correlated subgraphs, and the topology of  connecting network.  
Within the exhaustive spectral analysis for both the adjacency matrix and the normalized Laplacian matrix we identify the spectral properties which characterize the mesoscopic structure of sparse cyclic graphs and trees.  
The minimally connected nodes, clustering, and the average connectivity affect the central part of the spectrum. The number of distinct modules leads to an extra peak at the lower part of the Laplacian spectrum in cyclic graphs. Such a peak does not occur in the case of topologically distinct tree-subgraphs connected on a tree. Whereas the associated eigenvectors remain localized on the subgraphs both in trees and cyclic graphs. We also find a characteristic  pattern of periodic localization along the chains on the tree for the eigenvector components  associated with the largest  eigenvalue  $\lambda^L = 2$  of the Laplacian. Further differences between the cyclic modular graphs and trees are found by the statistics of random walks  return times and hitting patterns at nodes
on these graphs. The distribution of first return times averaged over all nodes exhibits a stretched exponential tail with the exponent $\sigma \approx 1/3$ for trees and $\sigma \approx 2/3 $ for cyclic graphs, which is independent on their mesoscopic and global structure.
\end{abstract}
\maketitle
\section{Introduction}
{\it Complex dynamical systems and network mesoscopic structure.}
In recent years a lot of attention has been devoted to the problem of representing the complex dynamical systems by  networks and investigating their structural and dynamical properties \cite {boccaletti2006}.  These networks often exhibit inhomogeneity at all scales, from the local level (individual nodes), to mesoscopic (groups of nodes) and global network level. The mesoscopic inhomogeneity of networks may be defined as topologically distinct groupings of nodes  in a range from few nodes to large modules, communities, or different interconnected sub-networks. These subgraphs play an important role in the network's complexity along the line from the local interactions to 
emergent global behavior, both in the structure and the function of networks \cite{boccaletti2006,tadic2007}.  Hence, the characteristic subgraphs can be defined not only topologically but also dynamically, and different subgraphs appear to characterize different functional networks. In particular, {\it communities} are often studied in social networks \cite{communities}, topological {\it modules} \cite{metabolic} and characteristic dynamical {\it motifs} \cite{motifs} are found in  genetic interactions and communication networks, whereas 
{\it paths}  and {\it trees} appear as relevant subgraphs in the networks representing  biochemical metabolic processes  and neural networks \cite{neural_nets}. \textit{Chains}, representing a special type of \textit{motifs}, have been observed in networks of words in books and the power grid networks, \cite{Costa2008}. In these examples the mesoscopic topology  is related to dynamics of the whole network. On the other side, we have {\it multi-networks} consisting of a few interconnected networks in which the internal structure and possibly also dynamics might be different \cite{multinetworks}. Then the interaction between such diverse networks leads to emergent global behavior, as for instance in the networks representing interacting eco-systems \cite{ecosystems}.

Understanding the mesoscopic structure of networks in both  topological and dynamical sense is, therefore, of paramount importance in the quantitative study of complex dynamical systems. Recently much attention was devoted to network's topological modularity, such as community structure \cite{communities,danon2006,Vito-dyn-centrality,newman2007,mmbtLNCS,arenas2006,donetti2004}, where a wide range of methods are designed to find the appropriate network partitioning. Mostly these methods use the  centrality measures (i.e., a topological \cite{danon2006} or a dynamical \cite{Vito-dyn-centrality} flow) based on the  maximal-flow-minimal-cut theorem \cite{cormen2001}. Further effective approaches for graph partitioning utilize the statistical methods of maximum-likelihood \cite{newman2007,mmbtLNCS}, occurrence of different time scales in the dynamic synchronization \cite{arenas2006} and eigenvector localization 
\cite{donetti2004} in mesoscopically inhomogeneous structures. In more formal approaches, the definitions of different mesoscopic structures in terms of {\it simplexes} and their combinations, {\it simplicial complexes}, are well known in the graph theory \cite{algebraic_topology_book}. This approach has been recently applied \cite{MilanLNCS} to scale-free (SF) graphs and some other real-world networks.

{\it Spectral analysis of networks.} Properties of the eigenvalues and eigenvectors of the adjacency matrix of a complex network and of other, e.g., Laplacian matrices related to the network structure, contain important information that interpolates between the network structure and dynamic processes on it. One of the well studied examples  is the synchronization of phase-coupled oscillators on networks \cite{boccaletti2006,arenas2006,mcgraw2008,Jurgen}, where the smallest eigenvalue of the Laplacian matrix corresponds to the fully synchronized state.   The synchronization between nodes belonging to better connected subgraphs (modules) occurs at somewhat smaller time scale \cite{arenas2006,albertNJP} corresponding to lowest nonzero eigenvalues of the Laplacian, and the positive/negative components of the corresponding eigenvectors are localized on these modules \cite{donetti2004,boccaletti2006}. The spreading of diseases \cite{eigen_cen} and random walks and navigated random walks  \cite{tadic2007,bt-arw01,nr-rwcn-04,ktr-njp07} are other type of the diffusive processes on networks which are related to the Laplacian spectra.

 Compared to the well known semicircular law for the random matrices \cite{random-matrices}, the spectra of binary and structured graphs have additional prominent features, which can be related to the graph structure \cite{farkas2001,dorogovtsev2003,goh2001,donetti2004,samukhin2007,mcgraw2008,jost2008}. Particularly, some of  the striking differences found in the scale-free graphs are  the appearance of the central peak or a 'triangular form' \cite{farkas2001} and the power-law tail \cite{geoff1988,dorogovtsev2003} in the spectral density of the adjacency matrix,  which is related to the node connectivity. In the classical paper Samukhin {\it et al.} \cite{samukhin2007} elucidated the role of  the minimally-connected nodes on  the Laplacian spectra of  trees and uncorrelated tree-like graphs. They derived analytical expression for the spectral density of the Laplacian matrix.  
Other topological features of the graph, particularly the finite clustering \cite{mcgraw2008} and the presence of modules \cite{donetti2004}, have been also found to affect the Laplacian spectra. Attempts  to classify the graphs according to their spectral features were presented recently \cite{jost2008}.

In this paper we study systematically the spectral properties of a large class of sparse networks with mesoscopic inhomogeneities.
The topology of these networks at all scales may lead to qualitative differences in the spectra both of the adjacency and Laplacian matrix.   Having well controlled structure of the networks by the model parameters, we are able to 
quantitatively relate the spectral properties of the networks to their structure.  As explained below, we identify different regions of the spectra in which certain  structural features are  mainly manifested.  We further explore these networks by simulating the random walk dynamics on them. We focus only on two properties of random walks: hitting patterns and first-return time distribution, which are closely related to graph structure and spectrum of the Laplacian.   In this way we would like to emphasise deeper interconnections between the structure and the dynamics of complex networks and their spectra, features which often remain fragmented in numerous studies of complex networks.

In Section\ \ref{sec:model} we represent model of growing networks with controlled number of modules and their internal structure. We then briefly study the spectral density of the adjacency matrix of modular networks in Section\ \ref{sec:adjasency}.
Section\ \ref{sec:laplace} is devoted to detailed  analysis of the spectra of the normalized Laplacian matrix, which is related to the diffusive dynamics on these networks. The simulations of random walks  on trees and on sparse modular graphs with minimal connectivity $M\geq 2$ is presented in Section\ \ref{sec:RW}. Finally, a short summary and the discussion of the results is given in Section\ \ref{sec:conclusions}.

\section{Growing Modular Networks \label{sec:model}}

We first present the model for growing networks with statistically defined modularity. It is based on the model for growing clustered scale-free graphs  
 originally introduced in Ref.\ \cite{tadic2001}. The preferential-attachment and preferential-rewiring during the graph growth leads to the correlated scale-free structure, which is statistically similar to the one in real WWW \cite{tadic2001}. Two parameters, $\alpha$ and $M$ as explained below, fully control the emergent structure.  
Here we generalize the model in a nontrivial manner by allowing that a new {\it module} starts growing with probability $P_0$. The added nodes are attached preferentially {\it within the currently growing module}, whereas the complementary rewiring process  is done between all existing nodes in the network.   The growth rules are explained in detail below.

At each  time step $t$ we add a new node $i$ and $M$ new links. With probability $P_{o}$ a new group (module) is started and  the current group index is assigned to the added node (first node belong to the first group). The group index plays a crucial role in linking of the node to the rest of the network. 
Note that each link is, in principle, directed, i.e., emanating from the {\it origin node} and pointing to the {\it target node}. For each link the target node, $k$, is always searched within the currently growing module (identified by its group index $g_k$). The target is selected preferentially according to its current number of {\it incoming} links $q_{in}(k,t)$. The probability $p_{in}(k,t)$ is normalized according to all possible choices at time $t$
\begin{equation}
p_{in}(k,t)=\frac{M\alpha+q_{in}(k,t)}{MN_{g_k}(t)\alpha+L_{g_{k}}(t)} \ . \label{pin}
\end{equation}
where $ N_{g_k}(t)$ and $L_{g_k}(t)$ stand for, respectively, the  number of nodes and links within the growing module $g_k$. The link $i\to k$ is fixed with the probability $\alpha$. If $\alpha <1$, there is a finite probability $1-\alpha$ that the link from the new added node $i\to k$ is cut (rewired) and a new origin node $n$ is searched from which the link $n\to k$ established and fixed. The new origin node $n$ is searched within all nodes in the network present at the moment $t$. The search is again preferential but according to the current number of outgoing links $q_{out}(n,t)$\cite{tadic2001}:  
\begin{equation}
 p_{out}(n,t)=\frac{M\alpha+q_{out}(n,t)}{MN(t)\alpha + L(t)} \ ,
\label{pout}  
\end{equation} 
where $N(t)=t$ and $L(t)\leq MN(t)$ are total number of nodes and links in the entire network at the moment $t$. Note that the number of added links is smaller than $M$ for the first few nodes in the modul until $M-1$ nodes are in the module. We are interested in sparse networks, for instance $M=2$, the second added node in a new module can have only one link pointing to the first node in that module. The second link is attempted once within the rewiring procedure the  probability $1-\alpha$, otherwise it is not added. 
It is also assumed that nodes have no in-coming or out-going links when they are added to the network, i.e., $q_{in}(i,i)$=$q_{out}(i,i)=0$. Some examples of the emergent modular graphs of size $N=10^3$ nodes are shown in Fig.\ \ref{fig-graphs4}. The networks which we consider throughout this paper are:

\begin{widetext}
\begin{figure*}[htb]
\begin{center}
\begin{tabular}{cc} 
\large{(a)}&\large{(b)}\\
\resizebox{16pc}{!}{\includegraphics{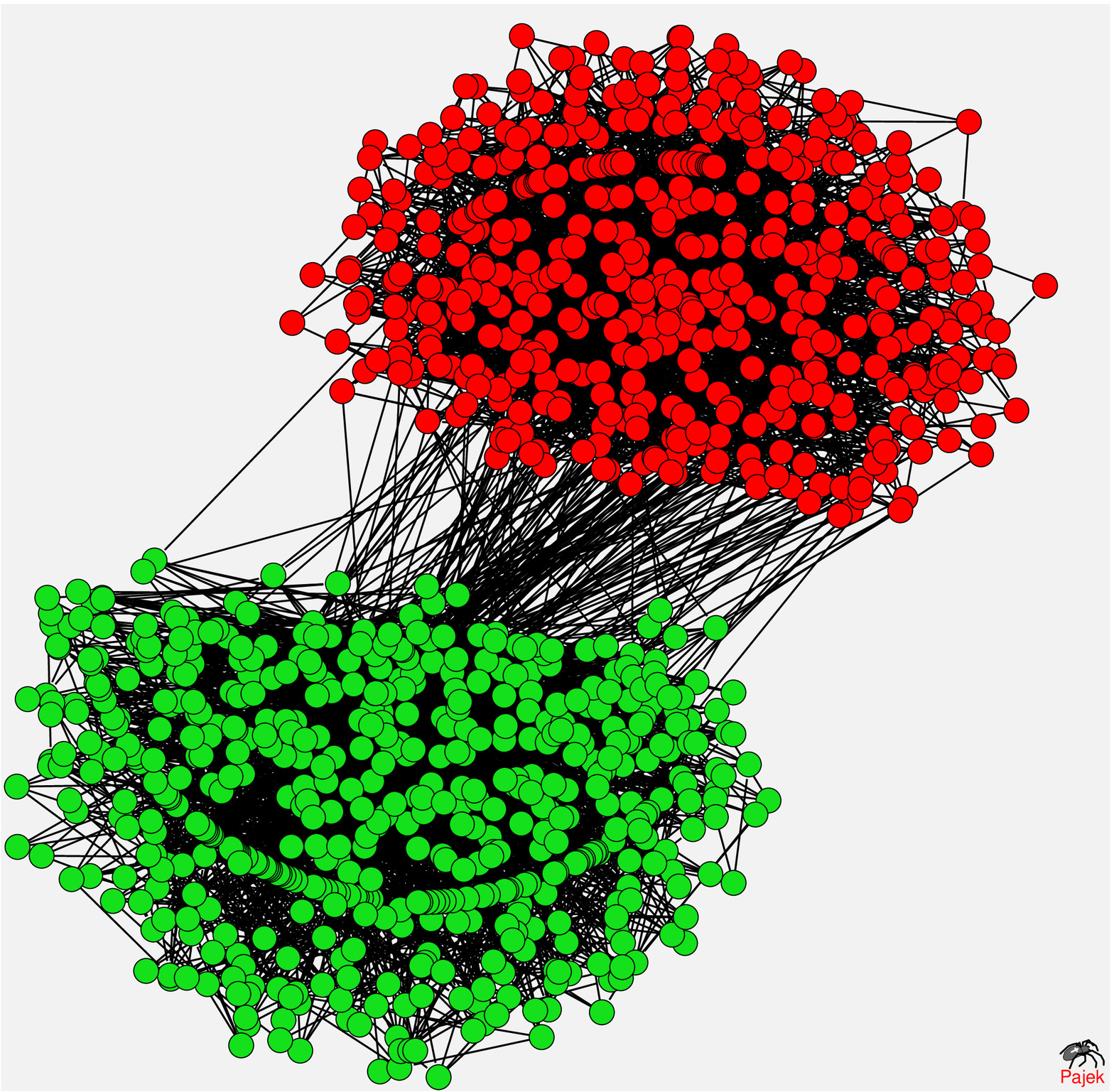}} &
\resizebox{16pc}{!}{\includegraphics{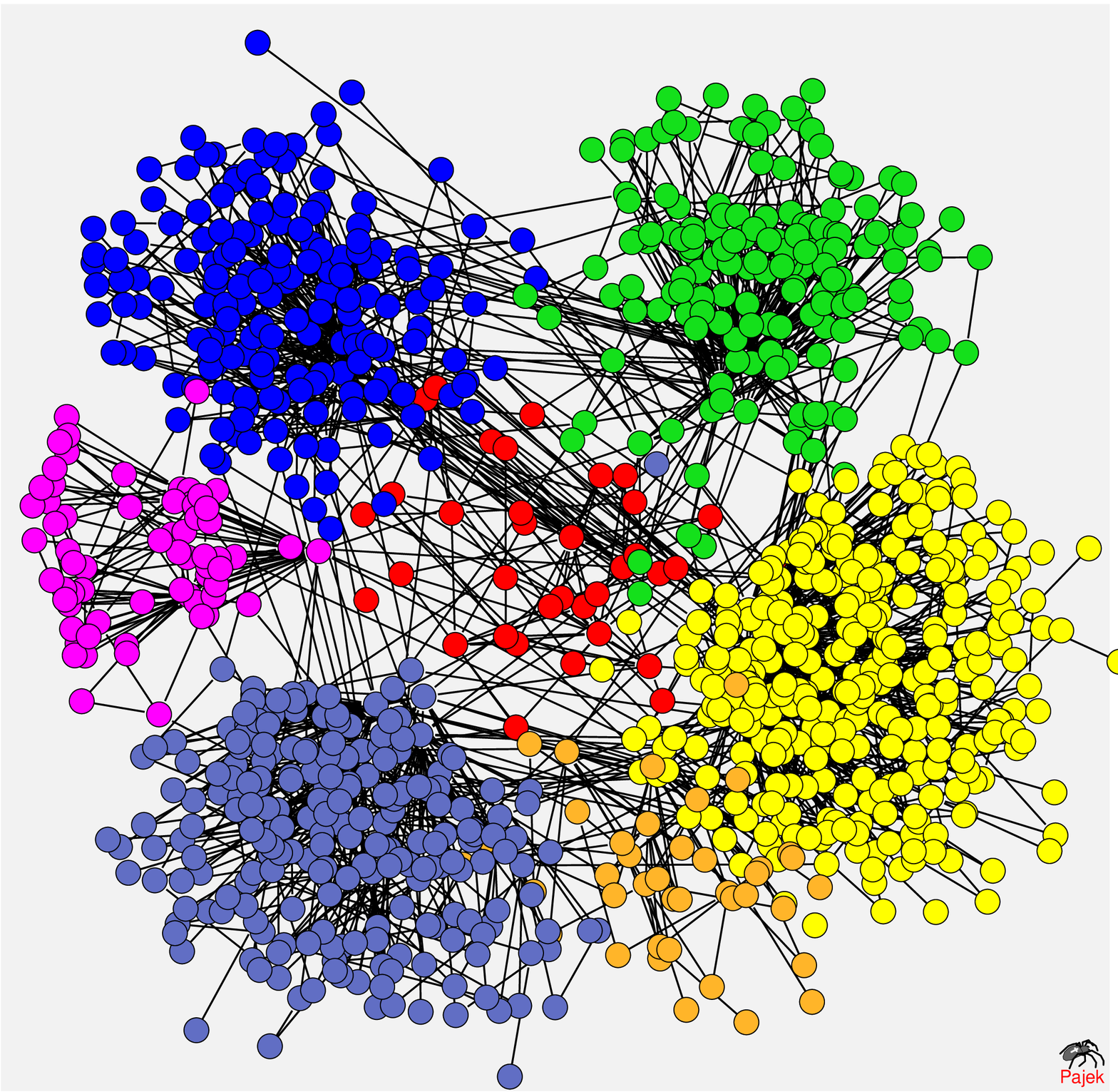}}\\
\resizebox{16pc}{!}{\includegraphics{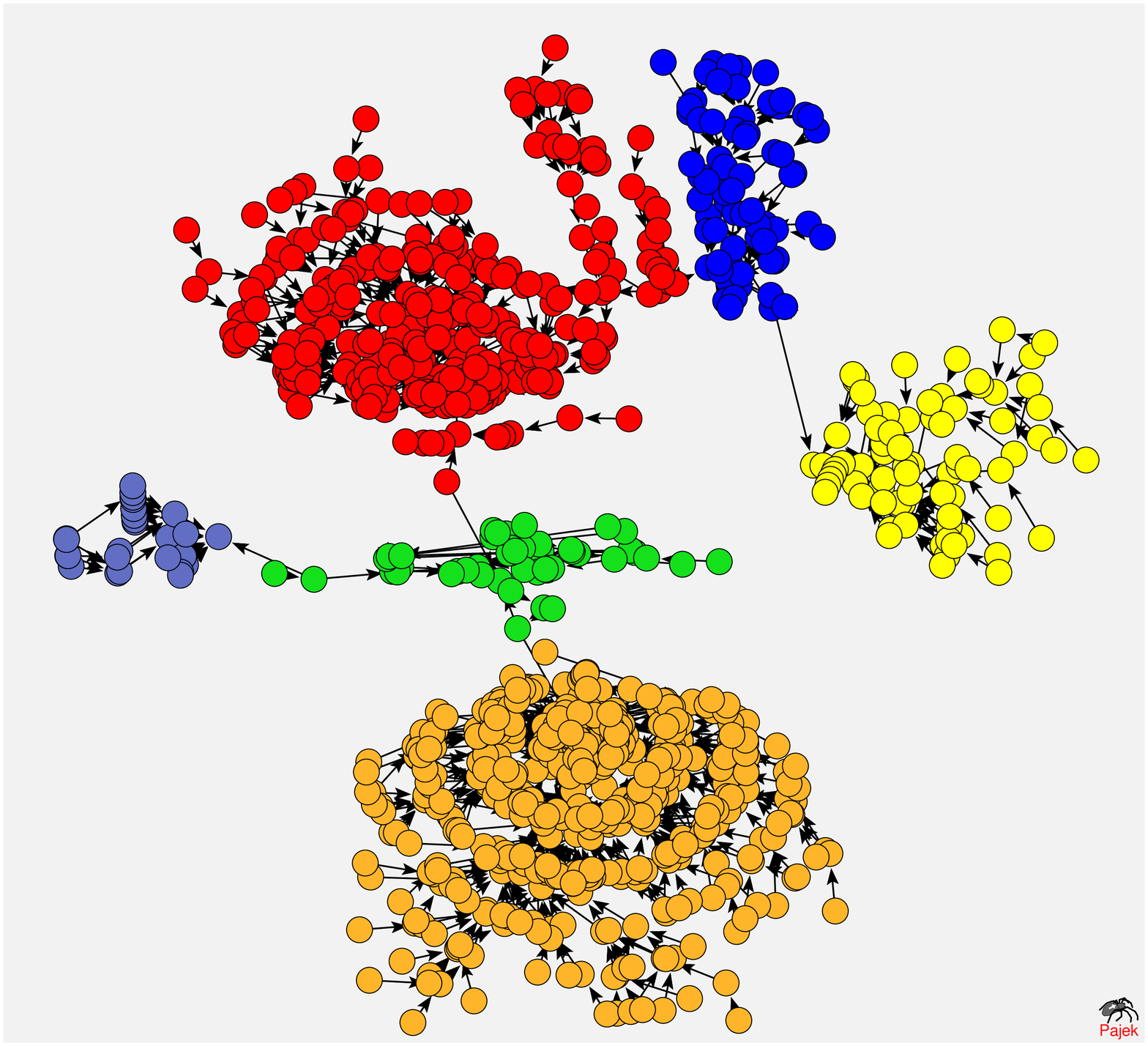}}&
\resizebox{16pc}{!}{\includegraphics{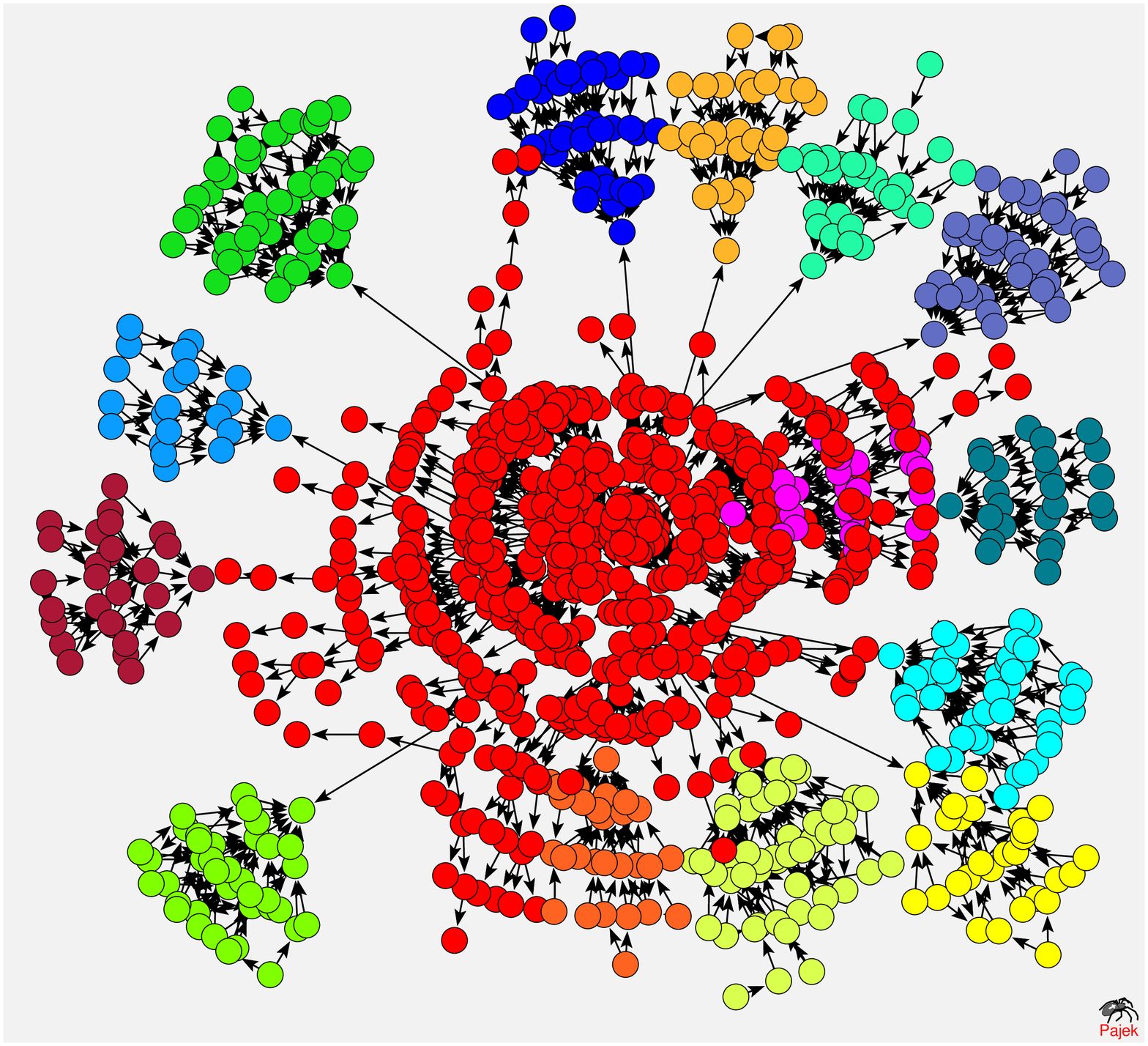}}\\
\large{(c)}&\large{(d)}\\
\end{tabular}
\end{center}
\caption{ (Color online) Examples of modular graphs with $N=1000$ nodes and $M \times N$ links grown from the model rules for different values of the control parameters: (a) $M=5$, $P_{o}=0.002$, $\alpha=0.9$;  (b) $M=2$, $P_{o}=0.006$, $\alpha=0.9$ (Net269); (c) $M=1$, $P_{o}=0.006$, $\alpha=1$ (Net161);  (d) Scale-free tree with attached modules. Each module contains between 20 - 50 nodes and its structure is determined by the growth rules with  $M=2$, $\alpha=0.9$. Color/gray scale of nodes indicates their group index.  }
\label{fig-graphs4}
\end{figure*} 
\end{widetext}
 
{\it Net269} (shown in Fig.\ \ref{fig-graphs4}b), is grown with direct implementation of the above rules with parameters $M=2$,  $P_0=0.006$, which gives $G\approx P_0N \geq 6$ distinct modules, and  $\alpha =0.9$, leading to $10\%$ links rewired.\\
{\it Net161}  (shown in Fig.\ \ref{fig-graphs4}c) is a scale-free tree  with tree subgraphs, which is grown with the same rules as above and taking  $M=1$, $G=6$ and $\alpha =1$ (no rewiring).

Two additional networks discussed in this appear are trees with attached modules of different structure. They are grown in the following way:\\
{\it Tree with SF-modules} (shown in Fig.\ \ref{fig-graphs4}d) starts growth as a scale-free tree, i.e., with $M=1$, $\alpha =1$ and preferential selection of the target node with the probability $p_{in}(k,t)= \frac{\alpha+q_{in}(k,t)}{N(t)\alpha +L(t)}$. A random integer $r$ in the interval $[20,50]$ is selected and at $r$th node a module of size $r$ is started to grow. The module rules are preferential linking and preferential rewiring {\it within the same module} with the parameters $M=2$ and $\alpha =0.9$.
Subsequently a new integer $r$ is selected and the  tree resumes to grow for the following $r$ steps, after which a module of the the same size is added and so on. Now the nodes in the modules are excluded as potential targets for the resumed tree growth. The relative size of the tree and modular structure can be controled in different ways, e.g., by the parameter $P_0$ as above. For the purpose of the present study we keep  full balance between the size of the tree and total number of nodes included in the modules.\\
{\it Tree with cliques} is grown as a random tree, i.e., target node $k$ is selected with probability $p_{in}(k,t)=1/N(t)$ from all nodes present at time $t$. With probability $P_0$ a clique of size $n$ is selected and attached  to a randomly selected node. Then the tree resumes growth and so on.

The
structural properties of these networks depend crucially on three control parameters: the average connectivity $M$, the probability of new group $P_0$, and the attractivity of node $\alpha$. By varying these parameters we control the internal structure of groups (modules) and the structure of the network connecting different modules. Here we explain the role of these parameters.
Note that for $P_0=0$ no different modules can appear and the model reduces to the case of the clustered scale-free graph of Ref.\ \cite{tadic2001} with a single giant component. In particular, for $M=1$ and $P_{o}=0$ and ${\alpha}<1$ the emergent structure is {\it clustered and correlated scale-free network}. For instance, the case $\alpha =1/4$ corresponds to the statistical properties measured in the WWW with two different scale-free distributions for in- and out-degree and nontrivial clustering and link correlations (disassortativity) \cite{tadic2001}.
On the other hand, for  $M=1$ and $P_{o}=0$, $\alpha =1$ a scale-free tree is grown with the power-law in-degree with the exponent $\tau=3$ exactly. 

Here we consider the case  $P_0 > 0$, which induces different modules to appear statistically. The number of distinct groups (modules) is given by $G\sim P_0N$. By varying the parameters $M$ and $\alpha$ appropriately, 
and implementing the linking rules as explained above with the probabilities given in Eqs.\ (\ref{pin}-\ref{pout}), we grow the modular graphs with $G$ connected modules of different topology. In particular for $\alpha <1$ the  scale-free clustered and correlated subgraphs appear (cf. Fig.\ \ref{fig-graphs4}a,b). Whereas for $\alpha =1$ the emergent structure is a tree of (scale-free) trees if $M=1$ (Fig.\ \ref{fig-graphs4}c). Another limiting case is obtained when $\alpha =1$ and $M\geq 2$, resulting in a scale-free tree connecting the unclustered uncorrelated scale-free subgraphs. In order to systematically explore the role of topology both of modules and connecting networks in the spectral and dynamical properties of sparse modular graphs,  we will study in parallel two network types shown in Fig.\ \ref{fig-graphs4}b and c, referred as Net269 and Net161, respectively.

\begin{figure}[htb]
\begin{center}
\begin{tabular}{cc} 
\includegraphics[width=9cm]{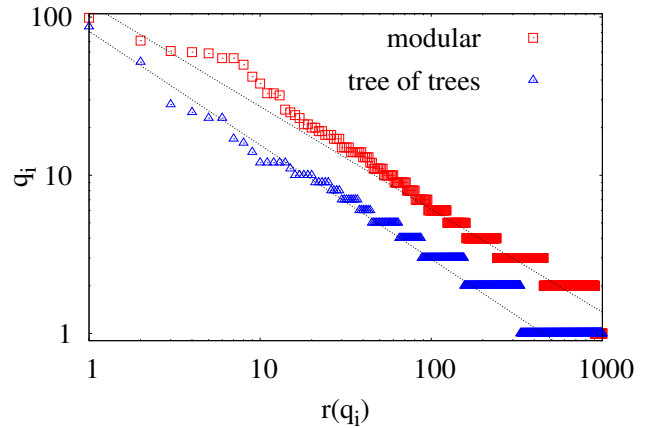} \\
\end{tabular}
\end{center}
\caption{Ranking distribution (Zipf's law) of nodes according to the total node degree $q_i$ for networks shown in Fig. \ref{fig-graphs4}b (modular) and \ref{fig-graphs4}c (tree of trees).}
\label{degree-ranking}
\end{figure}

Note that the growth rule as explained above lead to a directed graph with generally different connectivity patterns for in-coming and out-going links. Each module also tends to have a central node (local hub), through which it is connected with the rest of the network. The pattern of directed connections of the nodes within modules and the role of the connecting node can be nicely seen using the maximum-likelihood method for graph partitioning, as shown in our previous work \cite{mmbtLNCS}.
For the purpose of the present work, in this paper we analize {\it undirected binary graphs}, which have symmetric form of the adjacency matrix and the normalized Laplacian matrix. Therefore, the total degree of a node $q = q_{in} + q_{out}$ is considered as a relevant variable, for which we find a power-law distribution according to
 \begin{equation}
P(q)\sim q^{-\tau} . \label{pw}   
\end{equation}
In Fig.\ \ref{degree-ranking} we show the ranking  of nodes according to their degree for two networks, which are shown in Fig.\ \ref{fig-graphs4}b,c. The ranking distribution appear to be a broad (Zipf's law) with the exponent $\gamma$, which is related to the exponent in Eq.\ (\ref{pw}) with a general scaling relation
\begin{equation}
\tau=\frac{1}{\gamma}+1 \ . \label{cnn}
\end{equation}

The points in the flat parts of the curves at large connectivity represent the module hubs, which appear to have similar number of links. In the case of Net269 there are about six such nodes, whereas in the case of tree of trees two nodes, hubs of the largest subgraphs, are separated from four other hubs, and then the rest of nodes. The occurrence of local hubs changes the overall slope of the curve, compared to the networks without modules, where one finds analytically  $\gamma = 1/(1+\alpha)$ and thus $\tau=2+\alpha$ \cite{dorogovtsev2000,tadic2001}. Here we have approximately $\gamma\approx 0.65$ 
for modular network Net269, and  $\gamma \approx 0.72$ for tree of trees, Net161. According to Eq. (\ref{cnn}), $\tau \approx 2.5$ and $\tau \approx 2.4$, for these two networks, respectively, suggesting how the modularity affects the degree distribution.   

\section{Eigenvalue spectrum of Growing  Modular Networks\label{sec:adjasency}}

The sparse network of size $N$ is defined with an $N\times N$ adjacency matrix $\mathbf{A}$ with binary entries $A_{ij} = (1,0)$, representing the presence or the absence of a link between nodes $i$ and $j$. For the sparse binary networks the eigenvalue spectral density of the adjacency matrix is qualitatively different from the well known random matrix semi-circular law \cite{dorogovtsev2003}. Moreover, in a large number of studies it was found that  the eigenvalue spectra differ  for different classes of structured networks \cite{boccaletti2006,farkas2001,dorogovtsev2003,goh2001,donetti2004,samukhin2007,mcgraw2008,jost2008}. We study the spectral properties of the adjacency matrix ${\mathbf{A}}$ and the related Laplacian matrix ${\mathbf{L}}$ (see  Sec.\ \ref{sec:laplace}) of different networks with mesoscopic inhomogeneity using the complete solution of the eigenvalue problem: 
\begin{equation}
\mathbf{A}V^{A}_{i}=\lambda^{A}_{i}V^{A}_{i} \  . \label{evp}
\end{equation}
Here the set $\{\lambda^{A}_{i}\}$ denotes eigenvalues  and $\{V^{A}_{i}\}$  a set of the corresponding eigenvectors, $i=1,2 \cdots N$, of the adjacency matrix $\mathbf{A}$. For the modular networks grown with the algorithms in Section\ \ref{sec:model}  we focus on the effects that the network mesoscopic structure has on  (i) the spectral density, (ii) the  eigenvalues ranking, and (iii) the structure and localization of the eigenvectors.  

As stated above, we use the undirected networks. Thus the adjacency matrix is symmetric, which is compatible with the real eigenvalues  and the  orthonormal  basis of the eigenvector. We use the networks of the size $N=1000$ and solve  the eigenvalue problem  numerically. Particularly, we use the numerical routines in C from Numerical Recipes  \cite{nrc} for calculation of eigenvalues and eigenvectors of adjacency matrix with the precision  $10^{-6}$. The spectral densities are calculated with a large resolution, typically $\Delta\lambda=0.05$, and averaged over $500$  networks. 
 
\subsection{Change of the spectrum with network growth}

\begin{figure}[htb]
\begin{center}
\begin{tabular}{c} 
\includegraphics[width=8cm]{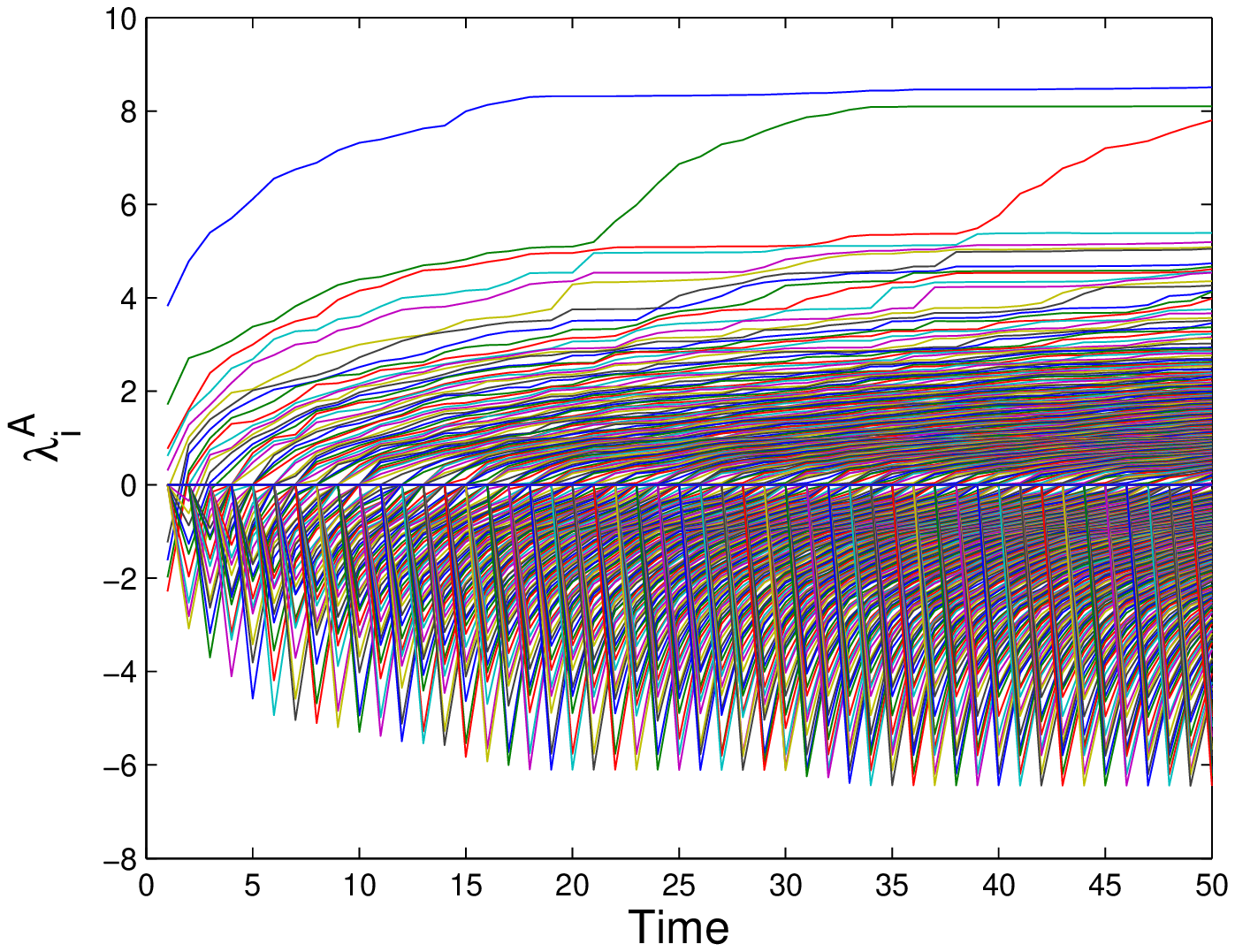} \\
\includegraphics[width=8cm]{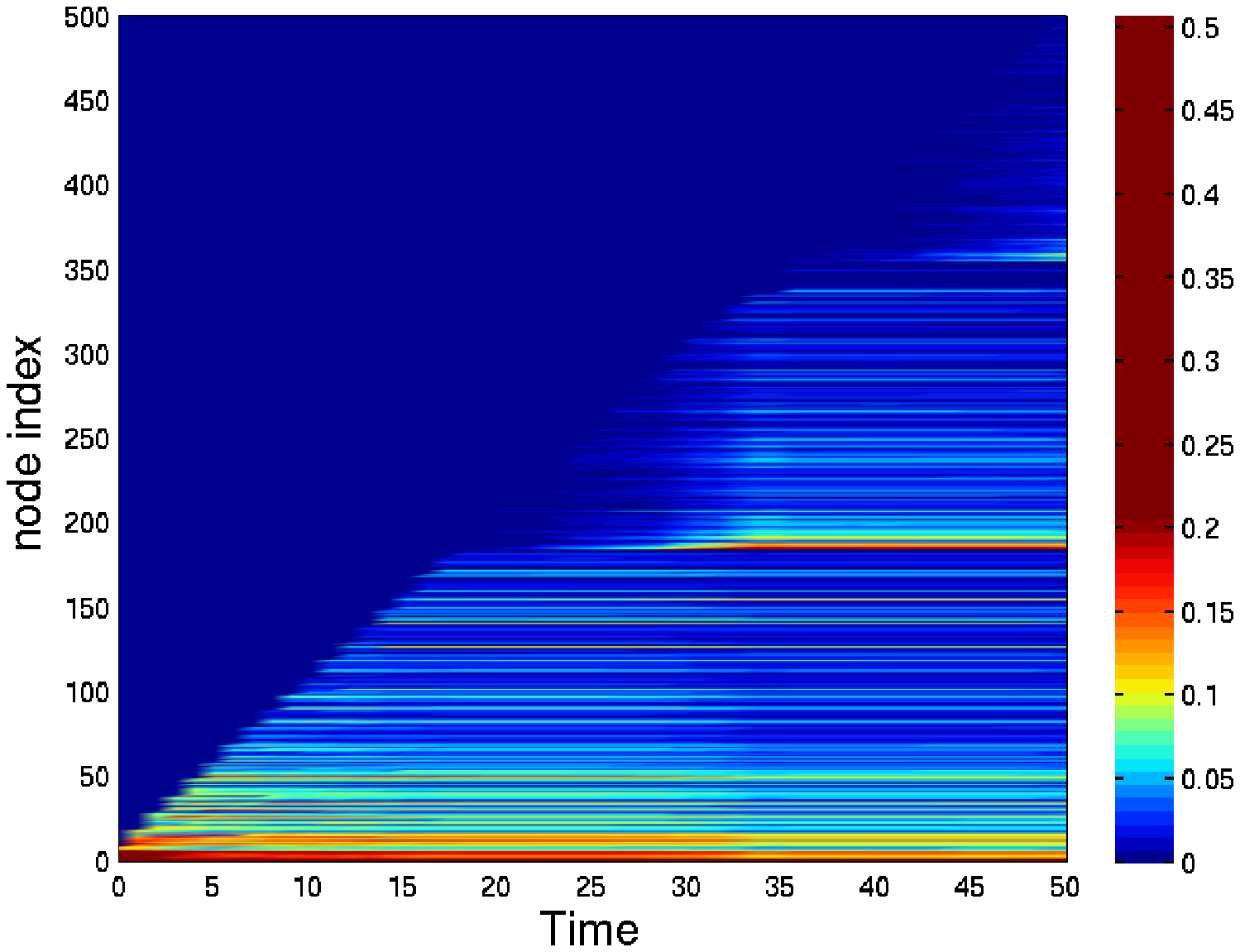} \\
\end{tabular}
\end{center}
\caption{(Color online.) Top: Evolution of all eigenvalues $\lambda ^A_i$ of the adjacency matrix with  growth of the modular network.  Bottom: Evolution of the components of the eigenvector associated with leading eigenvalue with network growth. Network parameters: $\alpha=0.9$, $M=2$ and $P_{o}=0.008$, permitting four modules within $N=500$ added nodes. Every tenth step is shown. Brighter/yellow-red color corresponds to larger centrality.}
\label{adj-growth}
\end{figure}

We first demonstrate how the growth of the modular networks affects their spectrum and the eigenvector components. In Fig.\ \ref{adj-growth} we show the eigenvalues of the growing network with the parameters selected such that   four modules are formed, i.e., at time step 1, 180, 338, and 357. Growth up to 500 added nodes was shown and the spectrum was computed every ten steps. 
As the network grows new eigenvalues appear, with the largest eigenvalue split from the bulk.  A remarkable feature of modular networks is that the additional eigenvalue splits from the rest of the spectrum 
when a new module starts growing.  In Fig,\ \ref{adj-growth} (top) three top lines corresponding to such eigenvalues are visible. The forth module is comparably  small. More detailed study of the eigenvalue spectral density will be discussed bellow. Here we show how the components of the eigenvector for the largest eigenvalue (eigenvector centrality) evolve in the same network.

The eigenvector centrality $x_{i}$ of a node $i$  satisfies the equation \cite{newman2003}
\begin{equation}
x_{i}=\frac{1}{\lambda ^A_{max}} \sum_{j=1}^{N}A_{ij}x_{j} \ , 
\label{ev-centrality}
\end{equation}
 hence, in view of the Eq.(\ref{ev-centrality}) and positivity of the centrality measures, it appears that different $x_i$ are the components of the eigenvector corresponding to the {\it largest} eigenvalue $\lambda ^A_{max}$ of the adjacency matrix $\mathbf{A}$. In the bottom panel of Fig.\ \ref{adj-growth} we show in a 3-dimensional color plot the evolution of the components corresponding to the largest eigenvalue of the growing network described above. In our network modules are interconnected ($\alpha <1$) what leads to the localization of the eigenvector on all nodes in the network. However, the largest component corresponds to the hub of the first module. When a new module is added to the network, 
the strongest component is eventually shared among the hubs of the two modules. During the growth of the module, however,  the centrality $x_i$ of the nodes in that module remains small until the module grows large enough (cf. Fig.\ \ref{adj-growth}).

\subsection{Spectral density of clustered modular networks}

 \begin{figure}[htb]
\begin{center}
\begin{tabular}{cc} 
\includegraphics[width=8cm]{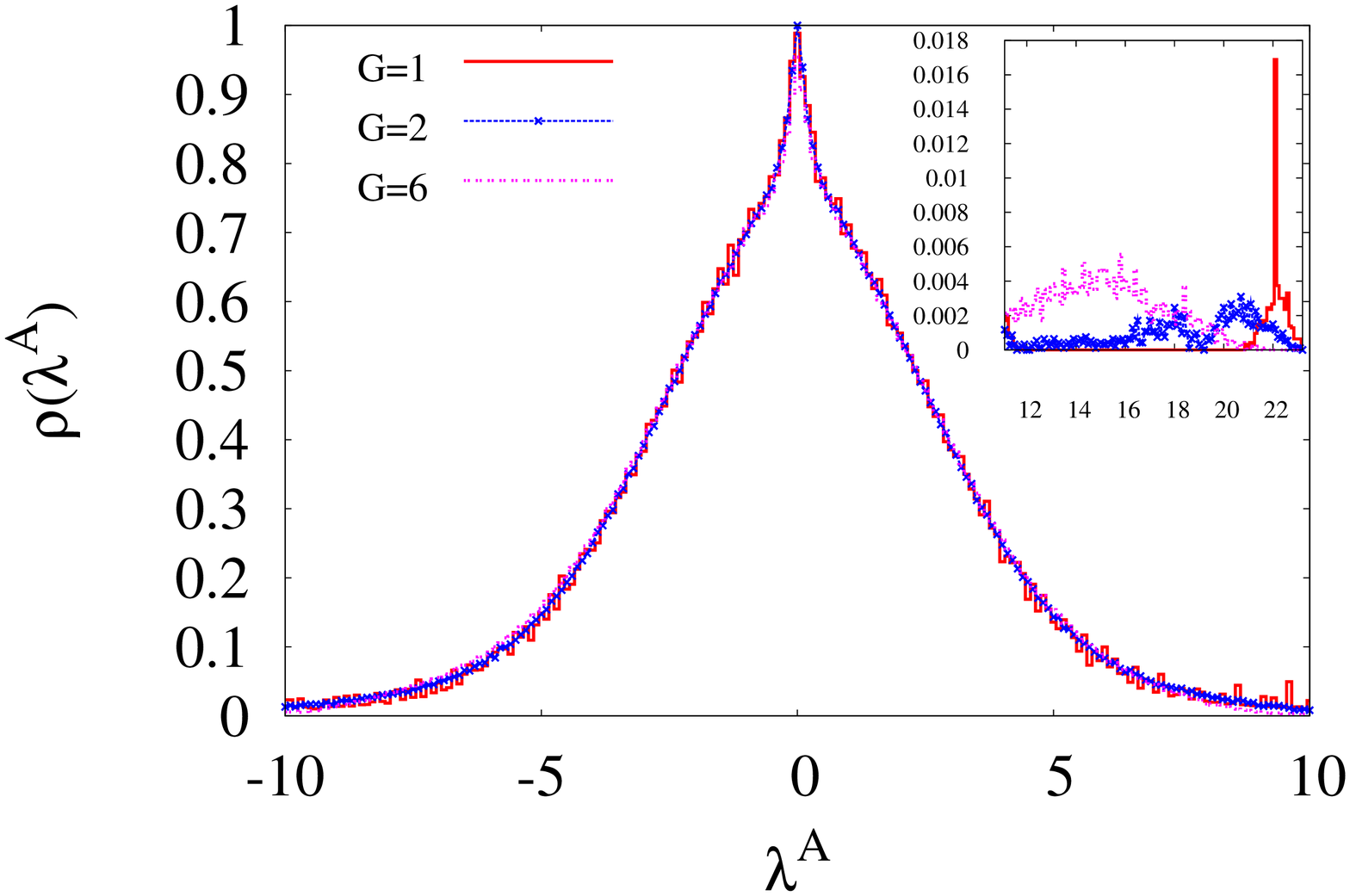} \\
\includegraphics[width=8cm]{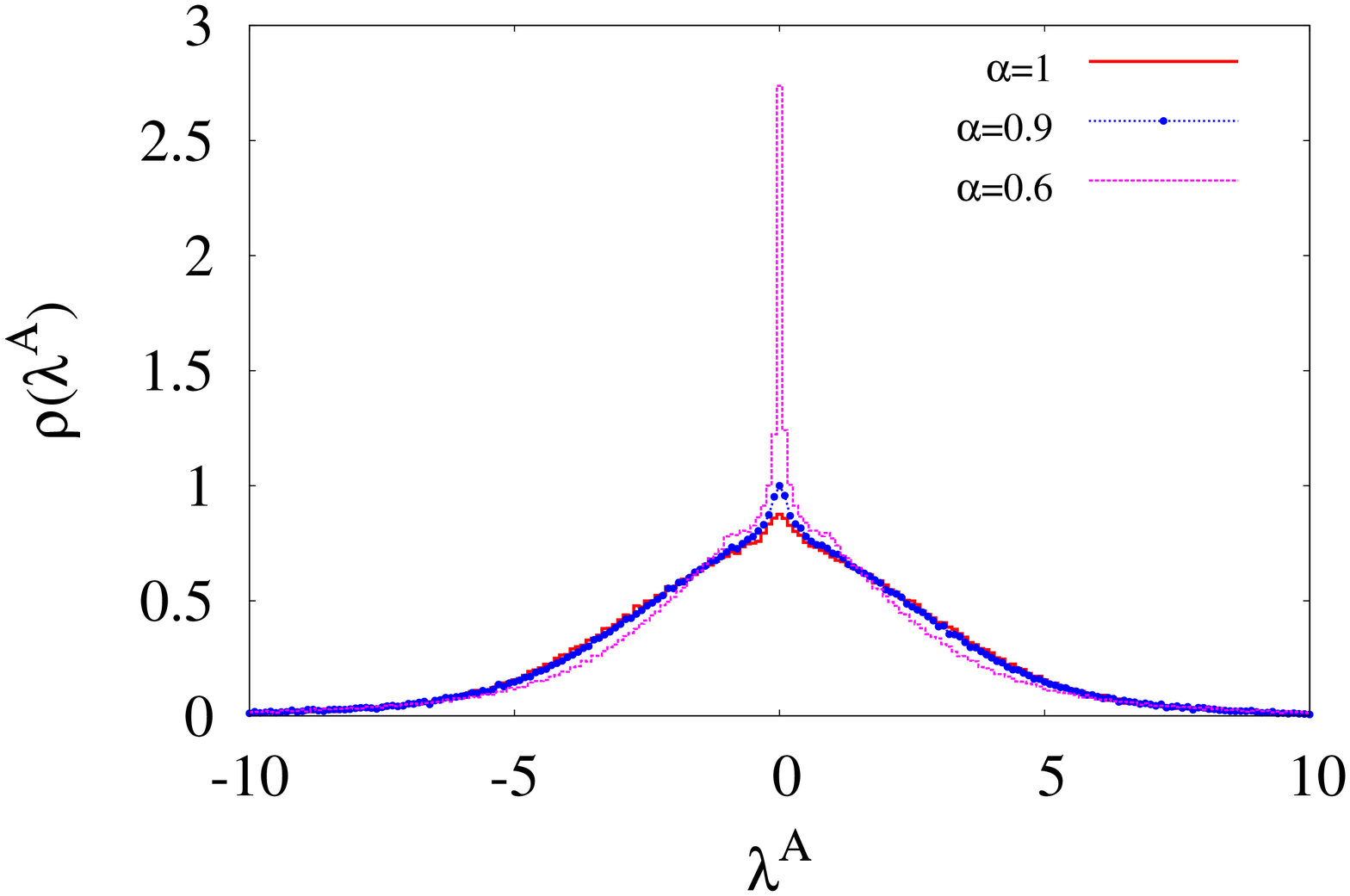}\\
\end{tabular}
\end{center}
\caption{(Color online.) Spectral density of the adjacency matrix for scale-free networks of size $N=1000$ and average connectivity $M=5$ for: (top)   $\alpha=0.9$ and varied modularity  $G=$1, 2, 6 modules. Inset:  Part of the spectrum with largest eigenvalues; (bottom)  fixed $G=2$ and varied  parameter of clustering $\alpha=$ 1, 0.9, 0.6, as indicated by color/type of line. Densities normalized to the maximum of the reference curve for $\alpha =0.9$, $G=2$.}
\label{adj_spectra_modular}
\end{figure} 
We investigate spectral densities of {\it undirected} networks grown with model presented in Section \ref{sec:model}. Our main focus is on the Laplacian spectra, studied in Section \ref{sec:laplace}. Here we briefly summarize the main features of the spectral density of the adjacency matrix of our scale-free graphs with the modularity and clustering (see also \cite{supplementaryPRE08}).
In the special case of our model when $P_0=0$ we have the scale-free networks without modularity. The spectral density of such networks, in particular unclustered and uncorrelated networks, which correspond to our case with $\alpha =1$, was investigated extensively \cite{farkas2001,goh2001,dorogovtsev2003,donetti2004,mcgraw2008,jost2008}. Specifically, it was shown that the spectral density  has a characteristic triangular shape and a tale, which is related to the power-law degree distribution \cite{dorogovtsev2003,geoff1988}. The largest eigenvalue is separated from the rest of the spectrum and its position scales with the largest connectivity $q_{max}$ of the hub, as $\sim \sqrt{q_{max}}$, \cite{farkas2001}. In Fig.\ \ref{adj_spectra_modular} (top) we show the spectral density of the unclustered ($\alpha =1$) scale-free networks with fixed average connectivity $M=5$ and varied number of modules. The case $G=1$ corresponds to  the case well studied in the literature \cite{farkas2001,goh2001,dorogovtsev2003,mcgraw2008,jost2008}. As the figure shows, for these type of the networks, the central part of the spectrum is not affected by the modularity ($G>1$) with fixed other parameters. The differences, however, appear in the area of the largest eigenvalue, as shown in the inset to Fig.\ \ref{adj_spectra_modular} (top). The number of different large eigenvalues increases with increased number of modules, as it was demonstrated in Fig.\ \ref{adj-growth}, which leads to broadening of the peak. At the same time, due to the fixed number of links $M\times N$, the largest connectivity is shared between several hubs of the modules, which leads to the shift of the peak towards lower values.

The internal structure of the modules is changed by varying the parameter $\alpha$. In particular, networks with different $\alpha$ have different degree distribution and for $\alpha < 1$ a finite clustering coefficient appears, which does not decay with the network size \cite{tadic2001}. In Fig.\ \ref{adj_spectra_modular} (bottom) we show the effects of increased clustering on the spectral density of networks with fixed average connectivity $M=5$ and fixed  number of modules $G=2$. The clustering coefficient for one network with $10\%$ of rewired links is, i.e., $\alpha =0.9$, is  $Cc=0.059$. For larger fraction of rewired links (decreasing $\alpha$) the clustering coefficient increases, for instance for the network with $\alpha=0.6$ we find   $Cc=0.164$. As shown in bottom panel of Fig. \ \ref{adj_spectra_modular}, the central part of the spectrum is affected by increased clustering of the network. In our model due to the preferential rewiring when $\alpha <1$, the number of triangles attached to hubs increases, while the peripheral nodes loose links. This contributes to the increase of the central peak and a decay of the spectral density away from the central area. The increased clustering also contributes to a characteristic shape around the central peak. Note that the random rewiring, which is often used to increase clustering in uncorrelated scale-free networks with large average connectivity, $M=20$ in Ref.\ \cite{mcgraw2008},  may  generate uncontrolable effects.

\section{Spectra of normalized Laplacian \label{sec:laplace}}
The Laplacian matrix $\mathbf{L}$ related to the adjacency matrix of the network $\mathbf{A}$ is usually defined as
\begin{equation}
L^{(1)}_{ij}=q_{i}\delta_{ij}-A_{ij} \  . \label{lap1}
\end{equation}  
For the dynamics of the random walks on networks other forms of the Laplacian matrices have been discussed in the literature \cite{samukhin2007,jost2008}. Generally, for a random or navigated walker \cite{guimera2002,tadic2005,agata2007} one can  define  the basic probability $p_{i\ell}$ for walker to jump from node $i\to \ell$ in a discrete time unit (one time step).   
Then the  probability $P_{ij}(n)$ that the walker starting at node $i$ arrives to node $j$ in $n$ steps is given by   
\begin{equation}
 P_{ij}(n)=\sum_{l_{1}\ldots l_{n-1}}p_{il_{1}}\ldots p_{l_{n-1}j} \ .  \label{prob1}
\end{equation}
Consequently, the  change of the transition probability $P_{ij}(n)$ in one time step can be written via 
\begin{eqnarray}
 &P_{ij}(n+1)&-P_{ij}(n)=\\\nonumber&
=&\sum_{l_{n}}[\sum_{l_{1}\ldots l_{n-1}}p_{il_{1}}\ldots p_{l_{n-1}l_{l_{n}}}](p_{l_{n}j}-\delta_{l_{n}j})\\\nonumber& 
\equiv&-\sum_{l_{n}}P_{i_{l_{n}}}(n)L_{l_{n}j} \ ,
\end{eqnarray}
which defines the components of the Laplacian matrix $L_{ij}$ in terms of the basic transition probability $p_{ij}$  of the walker. For the true random walk 
from node $i$ equal probability applies for all $q_i$ links, i.e.,   $p_{ij}=\frac{1}{q_{i}}$ when the link $A_{ij}$ is present. Thus the Laplacian matrix suitable for the true random walk on the network is given by  
 
\begin{equation}
L^{(2)}_{ij}=\delta_{ij}-\frac{1}{q_{i}}A_{ij} \  , \label{lap2}
\end{equation} 
and satisfies the conservation law for diffusion dynamics on graph \cite{jost2008}. 
We consider the symmetrical Laplacian 
\begin{equation}
L^{3}_{ij}=\delta_{ij}-\frac{1}{\sqrt{q_{i}q_{j}}}A_{ij} \  , \label{lap3}
\end{equation} 
which is a normalized version of the Laplacian for the random walks \cite{samukhin2007}. (It can be related with the transition probability chosen as  $p_{ij}=\frac{1}{\sqrt{q_{i}q_{j}}}$.)
The Laplacian matrix in Eq.\ (\ref{lap3}) has a limited spectrum in the range 
$\lambda _i^L \in [0,2]$ and an orthogonal set of the associated eigenvectors $V(\lambda _i^L)$, $i=1,2 \cdots N$, which makes it suitable for the numerical study. As already pointed out in Ref.\ \cite{samukhin2007}, the  
operators (\ref{lap2}) and (\ref{lap3}) are  connected by a diagonal similarity transformation $S_{ij}=\delta_{ij}\sqrt{q_{i}}$. Hence they have the same spectrum \cite {samukhin2007}.

\subsection{Spectral density of the normalized Laplacian of modular networks}
\begin{figure}[htb]
\begin{center}
\begin{tabular}{c}
\includegraphics[width=7.2cm]{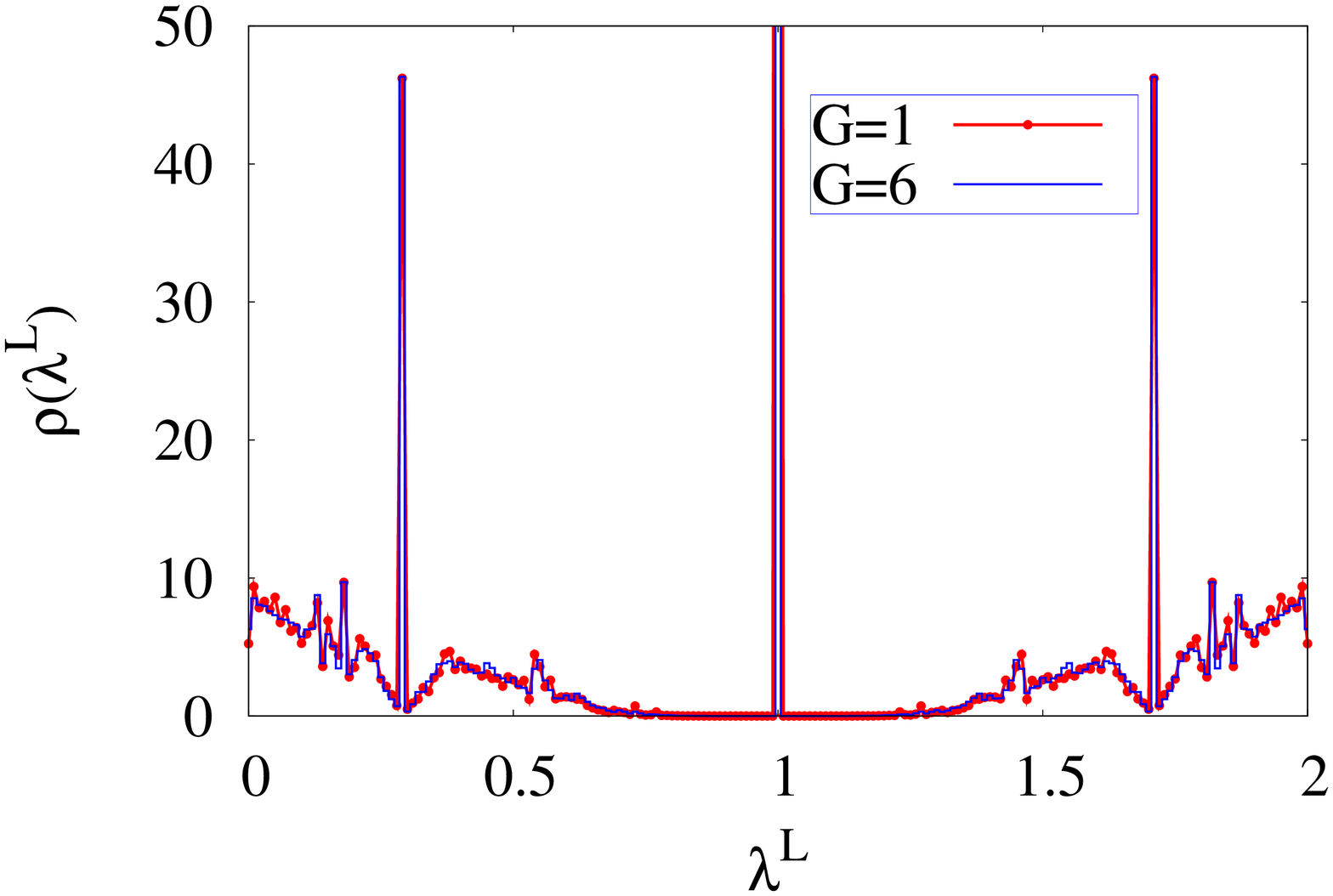}\\
\includegraphics[width=7.2cm]{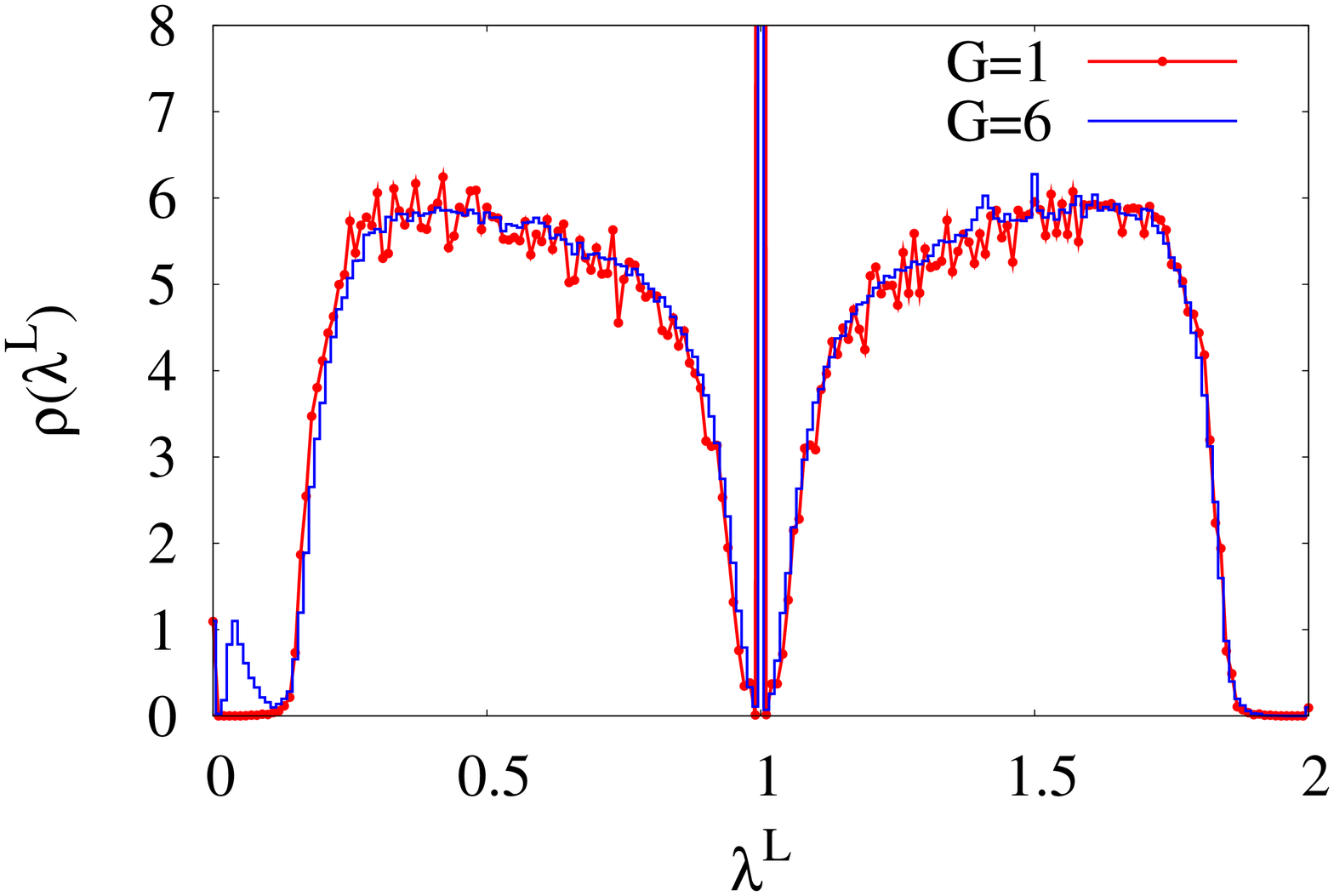}\\
\includegraphics[width=7.2cm]{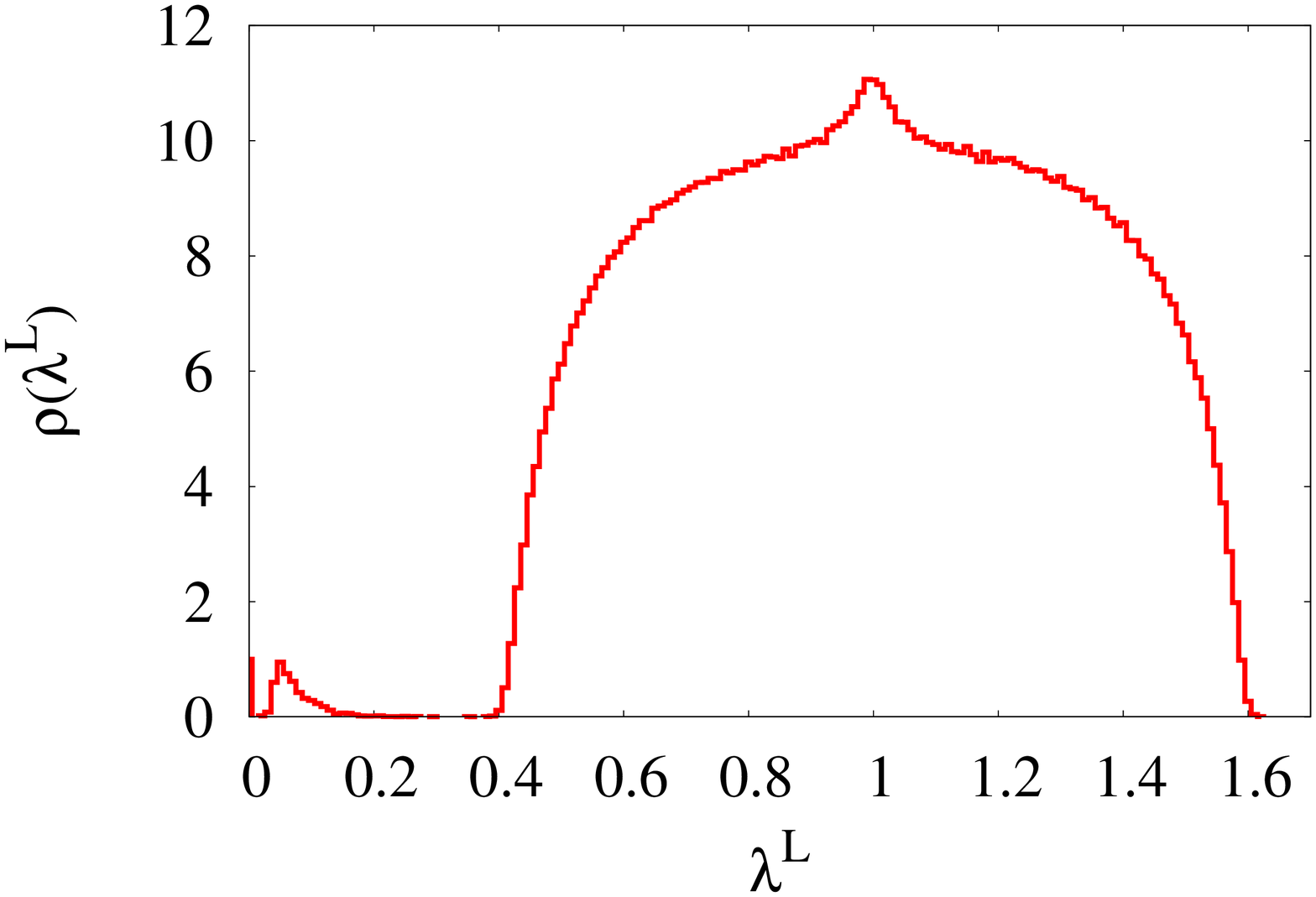}\\
\end{tabular}
\end{center}
\caption{(Color online.) Spectral density of the normalized Laplacian  for scale-free networks without modules, $G=1$, and with $G=6$ modules for scale-free trees  (top) and network with average connectivity $M=2$ and $\alpha=0.9$ (middle). Spectral density of the normalized Laplacian for scale-free network with $G=6$ modules and average connectivity $M=5$ and $\alpha=0.9$  (bottom). In each case the network size is $N=1000$ nodes and averaging is taken over 750 network samples.}
\label{laplacian_spectra}
\end{figure} 

\begin{figure}[thb]
\begin{center}
\begin{tabular}{c}
% \resizebox{14pc}{!}{ }
\includegraphics[width=8cm]{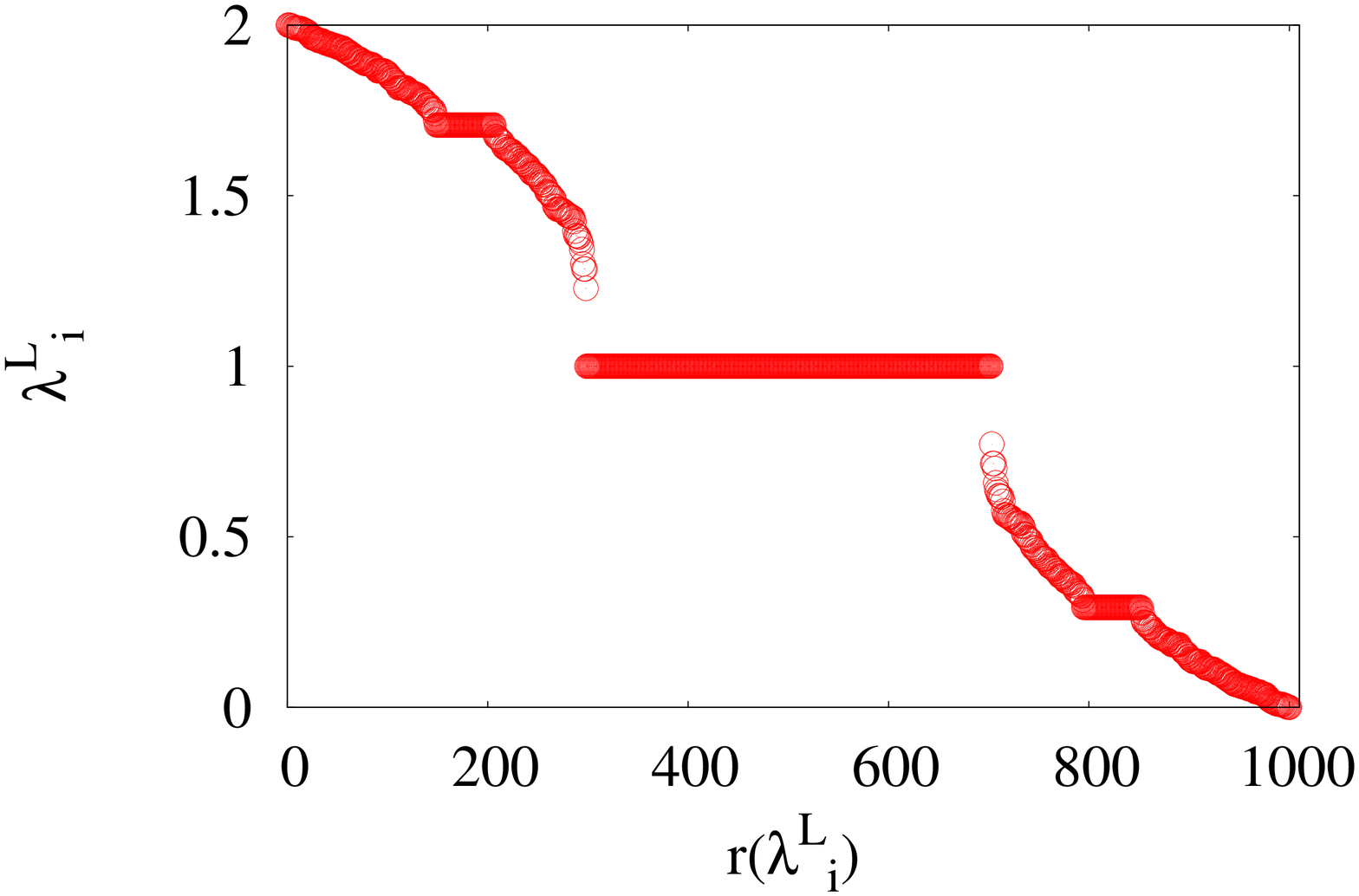}\\
\includegraphics[width=8cm]{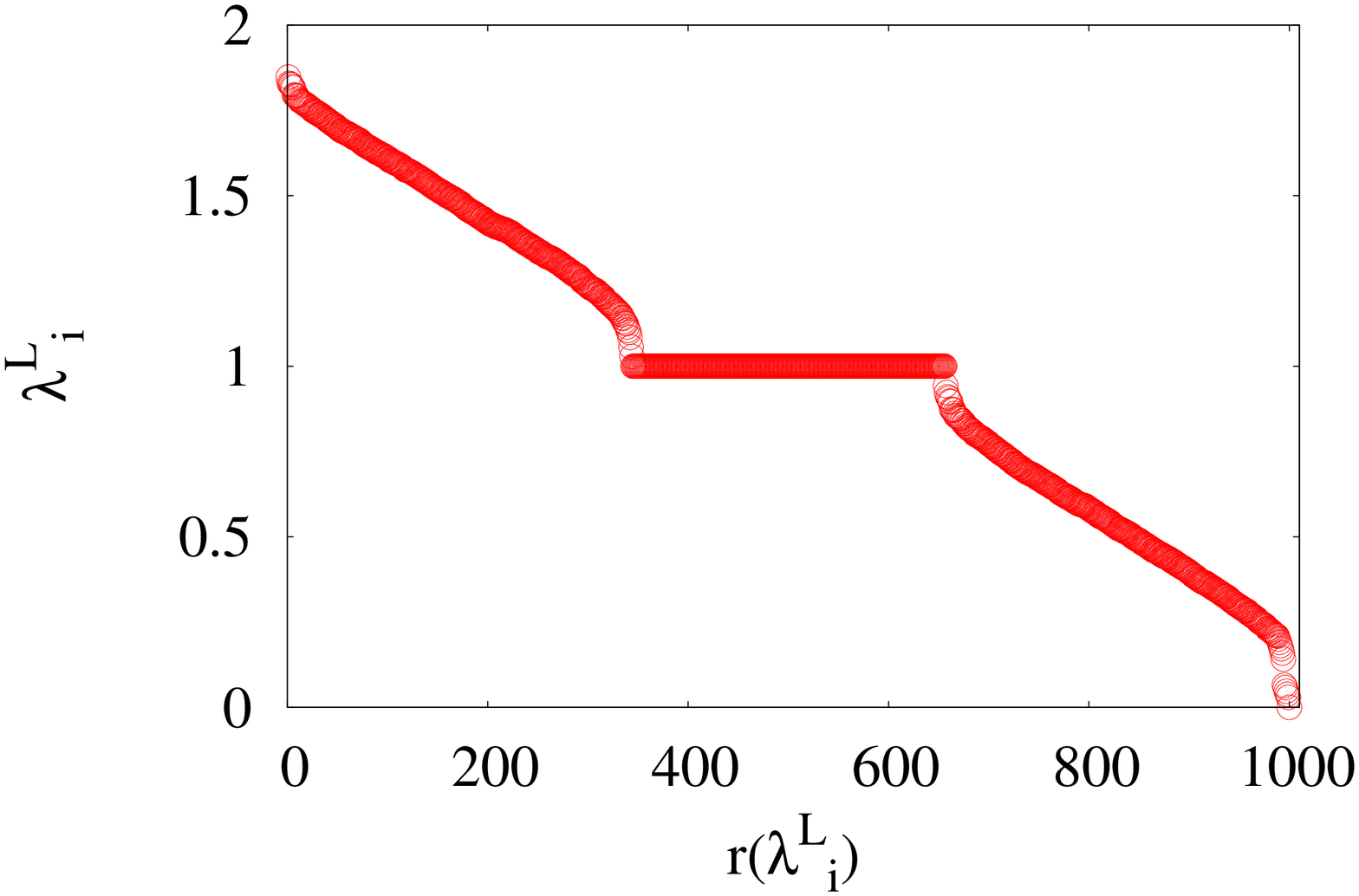}\\
\end{tabular}
\end{center}
\caption{Ranking of the eigenvalues for tree of trees, Net161 (top) and for modular network with $M=2$, Net269 (bottom).}
\label{ranking_tree_of_trees}
\end{figure} 
As mentioned above, the spectrum of the Laplacian (\ref{lap3}) is bounded  within the interval $[0,2]$, regardless of the size of network. The maximum value $\lambda_{max}^L=2$ is found only in bipartite graphs and trees. Whereas for monopartite graphs with cycles the maximum eigenvalue is shifted towards lower values, depending on the network structure. The main part of the spectrum is centered around unity. The minimum eigenvalue $\lambda _{min}^L =0$ always exists and it is non-degenerate if the graph consists of one connected component.  In the case of modular graphs, which we consider here, each of the modules tends to behave as an independent network with its own zero eigenvalue. 
This may  be manifested in the dynamics, for instance in the appearance of the  time scales for partial synchronization $t_i\sim 1/\lambda_i$ \cite{arenas2006,adg2008}, or in the confinement of the random walk inside the module, which affects the return times at small scale, as discussed later in Sec.\ \ref{sec:RW}.

Owing to the weak coupling between these subnetworks, we find one zero eigenvalue and a number of small eigenvalues $0 \lesssim \lambda $ corresponding to the number of topologically distinct modules. A typical spectral density of $L^{(3)}$ of an ensemble of networks with six modules and the average connectivity $M\geq 2$ shows the extra peak at small eigenvalues, as shown in Fig.\ \ref{laplacian_spectra} (middle) and (bottom) . In the sparse graphs and particularly in trees, the nodes with least number of links $q_m=1$ and $q_m=2$ play a special role in the form of the spectrum \cite{dorogovtsev2003} near the sharp peak in the adjacency matrix at $\lambda ^A=0$. Similar situation occurs at 
$\lambda ^L=1$, shown in Fig.\ \ref{laplacian_spectra} (top). Furthermore, in the case of trees we find continuous spectrum up to zero, although network like  Net161 in Fig.\ \ref{fig-graphs4}c has tree-like topological subgraphs. The topological modularity does not induce any new features of the Laplacian spectra in tree graphs. In the network Net269, Fig.\ \ref{fig-graphs4}b, we have 10\% of rewired links, which leave as much of the nodes with $q_m=1$. Consequently,  the central peak occurs, as shown in Fig.\ \ref{laplacian_spectra} (middle). The presence of cycles, however, leads to the two symmetrical peaks as well as the extra peak at small eigenvalues due to the presence of modules, Fig.\ \ref{laplacian_spectra} (middle). Comparison of the spectral densities in Figs.\ \ref{laplacian_spectra} (middle) and (bottom), suggests that the increase of the minimal connectivity of nodes while keeping the same number of topological modules, the central part of the spectrum approaches the one of a random binary graph (with disappearing central peak) and a gap opens between the lower and central part of the spectrum. 
The occurrence of the peak at lower part of the spectrum was noticed also in the earlier studies, for instance in highly connected network with $M=20$ in \cite{mcgraw2008}, where rewiring of a large number of links per node eventually leads to both increased clustering and probability of a modular structure. In our model, on the contrary, it is clear that the peak at the lowest part of the spectrum is indeed related to the topologically distinct modules in cyclic networks even if the networks are very sparse, i.e., $M= 2$  and $M=5$, as shown in Fig.\ \ref{laplacian_spectra}(middle) and (bottom). The width of the gap increases with the average connectivity. Also, the area under the small peak, compared to the central part of the spectrum  increases with the number of distinct modules. For instance, for $M=2$ we find (see Supplementary material \cite{supplementaryPRE08}) the relative weight of the small peak increases from $0.144\%$ at $G=2$ to $1.53\%$ at $G=16$ modules.

The spectral densities in Fig.\ \ref{laplacian_spectra} are obtained with ensemble average over many networks grown with using the same parameters $P_0, \alpha, M$. For the individual network realization, particularly the Net161 and Net269 shown in Fig.\ \ref{fig-graphs4}b,c, the eigenvalues are shown in the ranking order in Fig.\ \ref{ranking_tree_of_trees}.   Within the numerical precision, the central plateau in the ranking distribution corresponds to the sharp central peak in the spectral density of an ensemble. It is also clear that the cyclic Net269 has six lowest eigenvalues (lower right corner) separated from the rest of the spectrum, and the largest eigenvalue lies below 2. In the tree network Net161, however, the spectrum approaches both ends continuously. Additional plateau is found at the eigenvalue $\lambda^{L}=1.707107$ and symmetrically at $\lambda^{L}=0.292893$, corresponding to the side peaks in the spectral density (see Fig. \ \ref{laplacian_spectra}, top).

\begin{figure}[thb]
\begin{center}
\begin{tabular}{c}
\includegraphics[width=7cm]{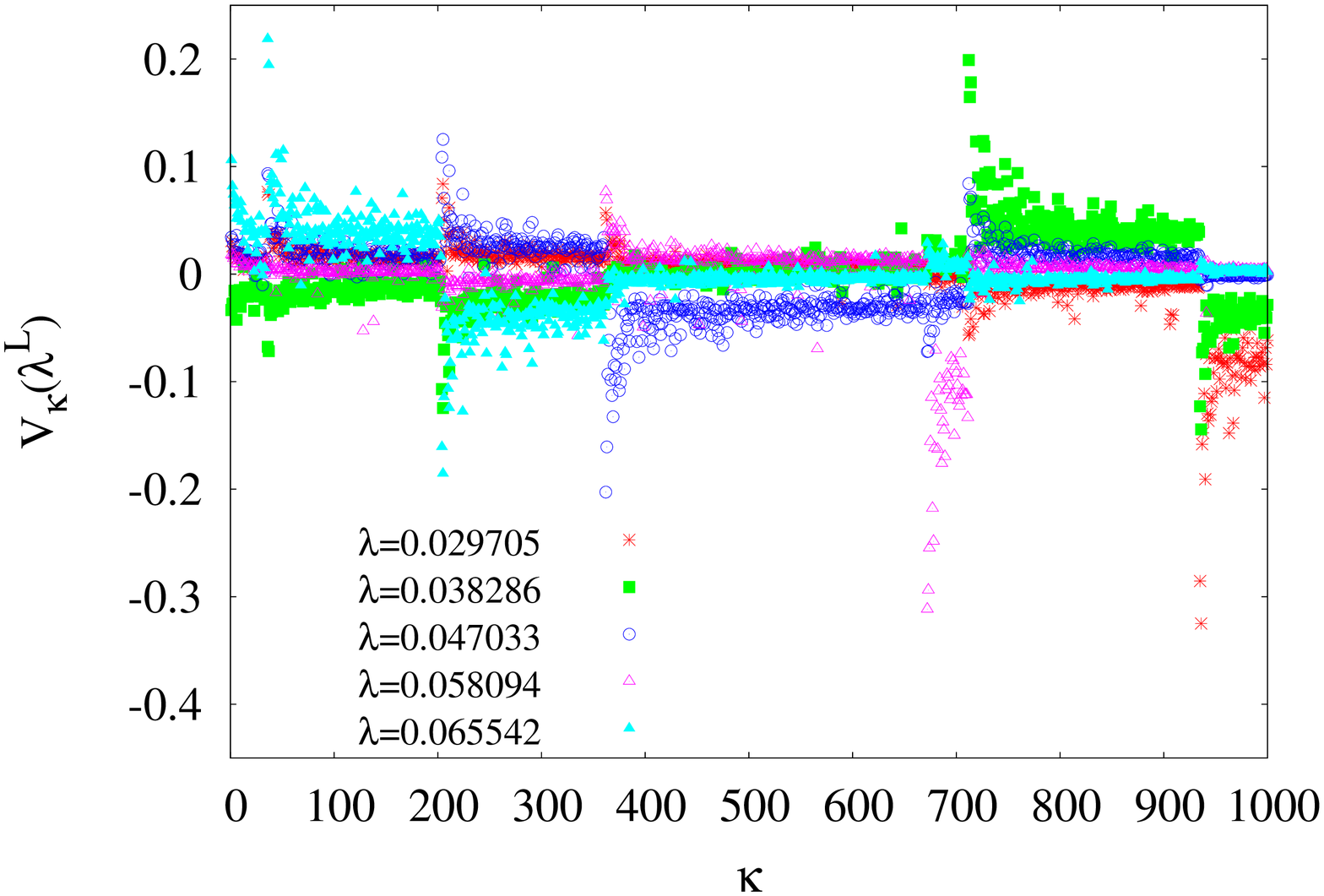}\\
\includegraphics[width=7cm]{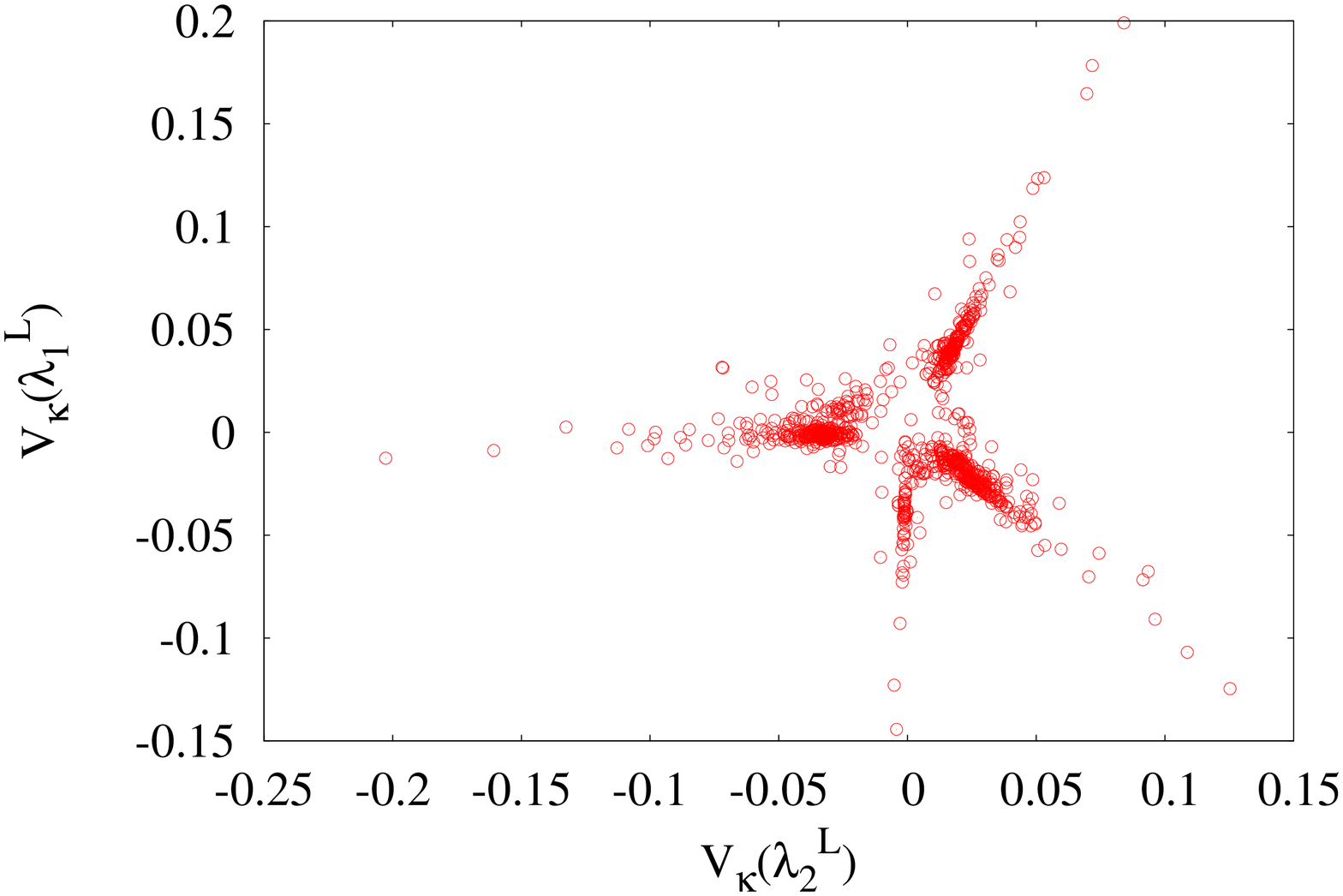}\\
\includegraphics[width=7cm]{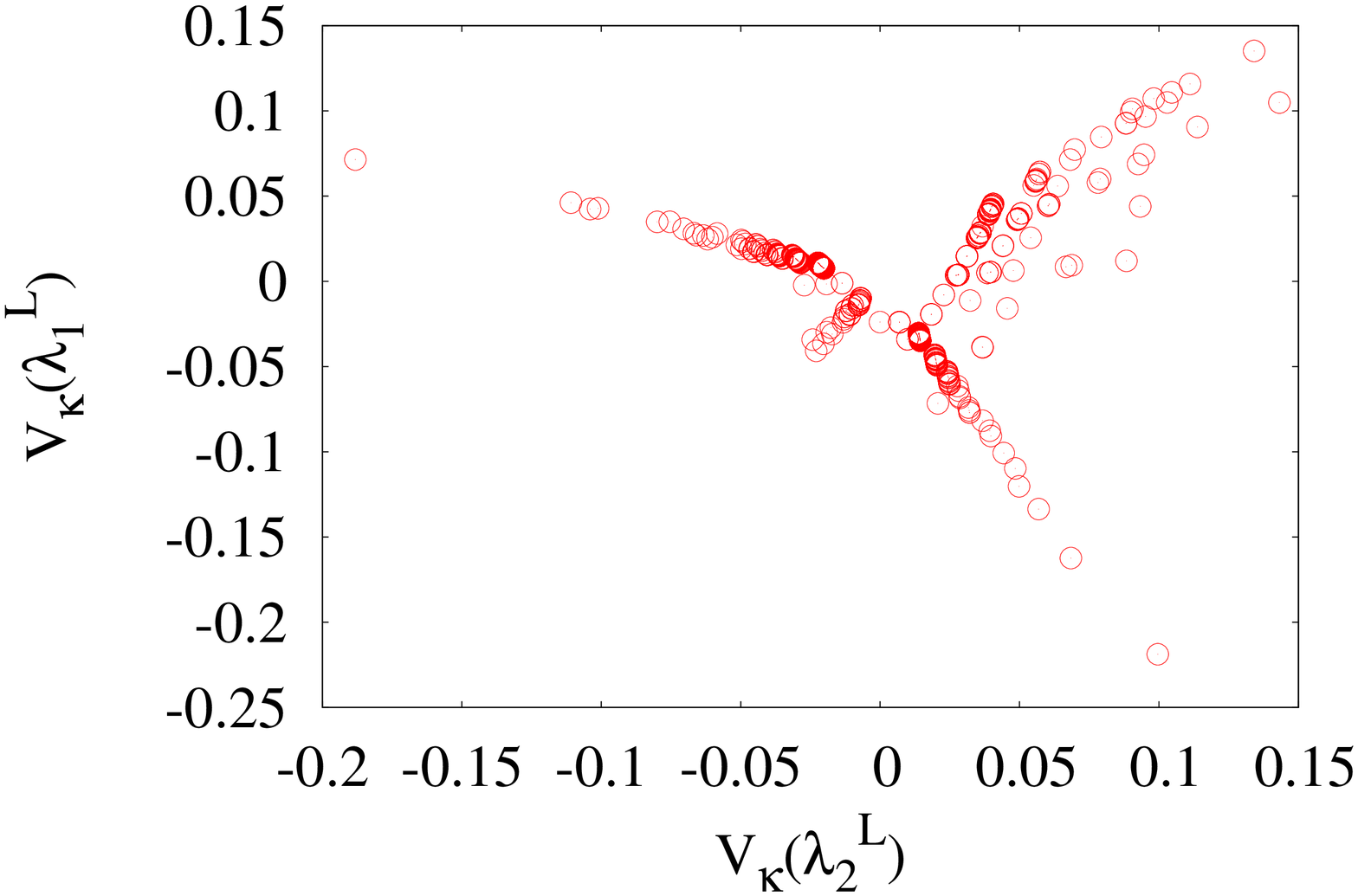}\\
\end{tabular}
\end{center}
\caption{(Color online.) Top: Eigenvector components, indicated by five colors/gray scale, for five lowest non-zero eigenvalues of the normalized Laplacian for the network Net269. Middle: Scatter plot of the eigenvector components for the eigenvalues $\lambda_{1}^{L}=0.047033$ and $\lambda_{2}^{L}=0.038286$ for the network Net269. Bottom: Scatter plot of the eigenvector components for the eigenvalues $\lambda_{1}^{L}=0.00108$ and $\lambda_{2}^{L}=0.000335$ for the tree of trees, Net161.}
\label{vec_lap}
\end{figure}

\subsection{Structure and Localization of the Eigenvectors}
Another prominent feature of the eigenvalue problem of the Laplacian matrix  
$\mathbf{L}^{(3)}$ is revealed by the structure and the localization of the components of the eigenvectors. For each eigenvalue $\lambda _i^L$, $i=1,2, \cdots, N$, we have an associated eigenvector $V(\lambda _i^L)$ with the components $V_\kappa$, $\kappa =1,2, \cdots N$. A {\it localization} implies that the 
 {\it nonzero components}  $V_\kappa \neq 0$  {\it of the eigenvector  coincide with a particular set of geometrically distinguished nodes on the network}.
Specifically, for the case of the cyclic graph, Net269, the eigenvectors associated with the lowest nonzero eigenvalues appear to be well localized on the network modules, as shown in Fig. \ref{vec_lap}(top). The origin of such localization of the eigenvectors corresponding to the lowest eigenvalues has been discussed in the literature \cite{donetti2004}, and it is related to the property of the Laplacian. The eigenvector corresponding to the trivial eigenvalue $\lambda ^L=0$ for the connected network has all positive components and $\kappa^{th}$  component is proportional to $\sqrt{q_{\kappa}}$. When networks consists of $G$ disconnected subgraphs, each of $G$ eigenvectors with $\lambda^{L}=0$ has non-zero components only for nodes within one module. If the subgraphs are not fully disconnected, but instead,  few links exist between them, the degeneration of the zero eigenvalue disappears, leaving only one trivial eigenvector and $G-1$ approximately linear combinations of the eigenvectors of the modules. 
For the orthogonality reasons, these linear combinations have components of both signs, as opposed to all positive components of the $V(\lambda ^L=0)$ vector.  In the case of well separated modules, the components corresponding to one subgraph appear to have the same sign, as shown in  Fig. \ref{vec_lap} for the case of Net269. The more links between subgraphs exist, the distinction between modules appears fuzzier. 
The structure is also seen in the {\it scatter plots} in Fig.\ \ref{vec_lap}(middle and bottom) where the eigenvector components belonging to two small eigenvalues are plotted against each other. In this projection each point corresponds to the index of one node on the network. 
The separated branches along the lines $y=\pm ax$ contain the indexes of the nodes belonging to different modules. A similar feature occurs in the case of tree of trees Net161, shown in Fig.\ \ref{vec_lap}(bottom). The occurrence of such patterns  related to the network modules is a direct consequence of the localization of the eigenvectors. Consequently, in the absence of the appropriate localization, for instance in the center of the spectrum, $\lambda ^L =1$, the corresponding scatter plot does not exhibit any pattern (not shown).

\begin{figure}[htb]
\begin{center}
\begin{tabular}{cc} 
\includegraphics[width=7.4cm]{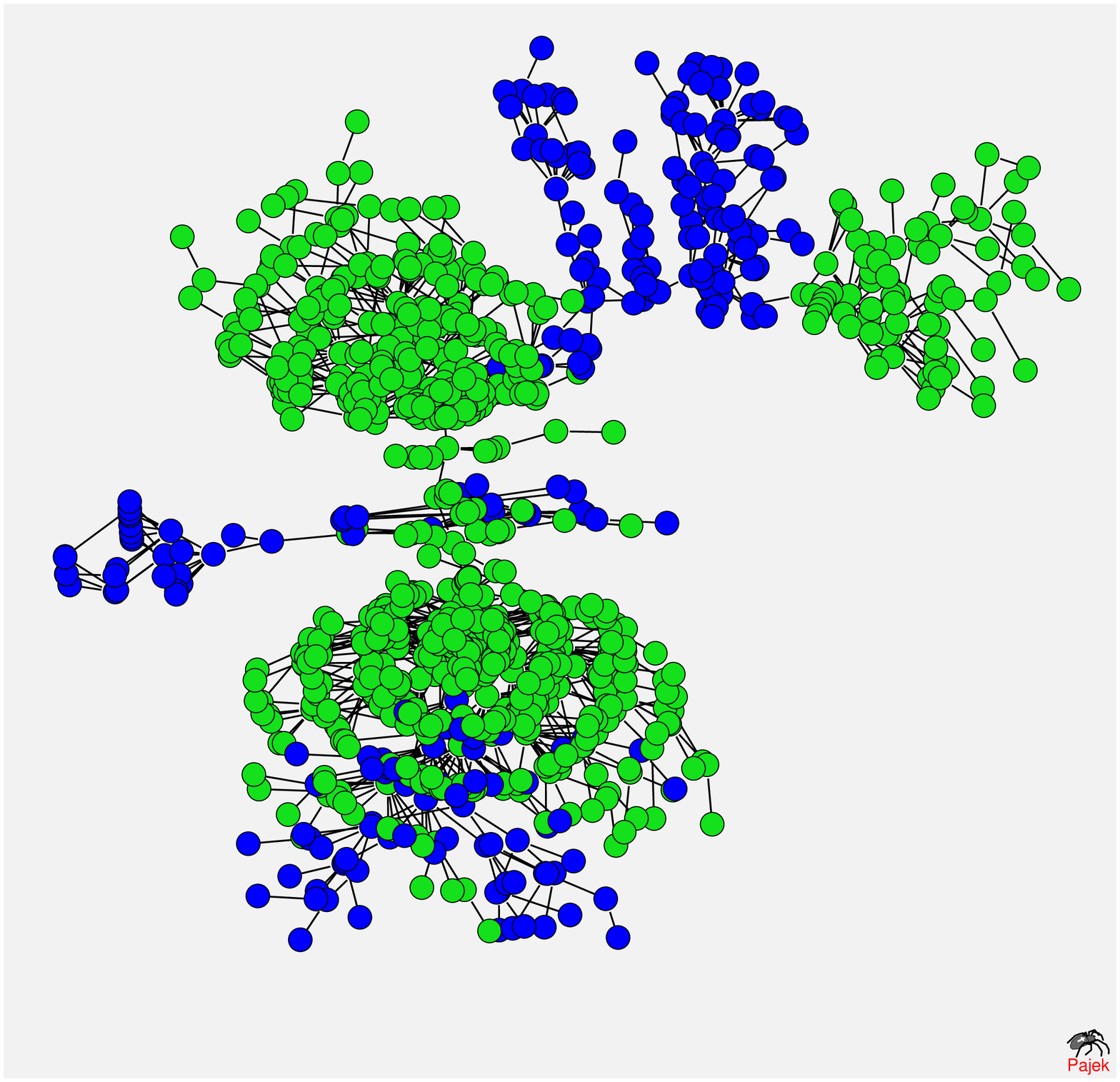}\\
\includegraphics[width=7.4cm]{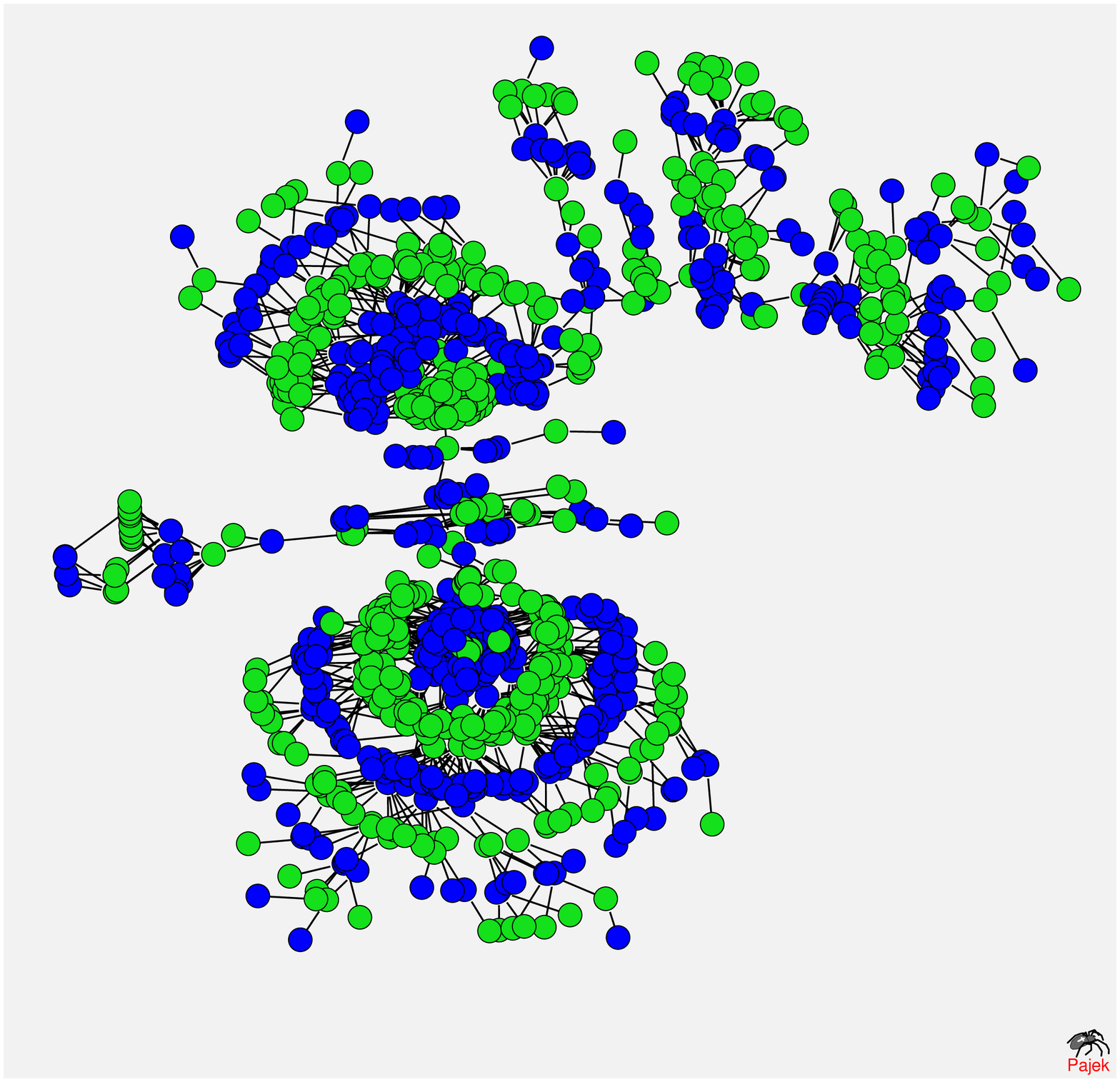}\\
\end{tabular}
\end{center}
\caption{(Color online.) Localization of the eigenvectors belonging to  (top) small eigenvalue $\lambda_{2}^{L}=0.004623$, and  (bottom)  largest eigenvalue $\lambda_{max}^{L}=2$ of the Laplacian on the tree of trees network, Net161. Dark (blue)/gray (green) color indicate positive/negative values of the eigenvector components.}
\label{ev-localization}
\end{figure}

It is interesting to note that in the case of our tree of trees, Net161, the localization of the eigenvectors for small $\lambda^L$ can be also observed, shown in Fig.\ \ref{ev-localization}a in "real space", although the separation of that part of the spectra is absent for trees, as discussed above. In addition, we find that the eigenvector associated to the largest eigenvalue on trees $\lambda _{max}^L=2$, also shows a characteristic pattern of localization with a succession of the positive-negative components along the network chains. The situation is shown in Fig.\ \ref{ev-localization}b. 
In fact, the regularity in the localization pattern indicates the bi-partitivity of the tree graph, which is associated with the $\lambda_{max}^L=2$. We find it interesting, that  the two partitions within the spectral analysis are not seen as different 'communities' at lower part of the spectrum, where rather the localization on the subtrees occurs, as shown in Fig.\ \ref{ev-localization}a. 

\begin{figure}[htb]
\begin{center}
\begin{tabular}{c} 
\includegraphics[width=7.2cm]{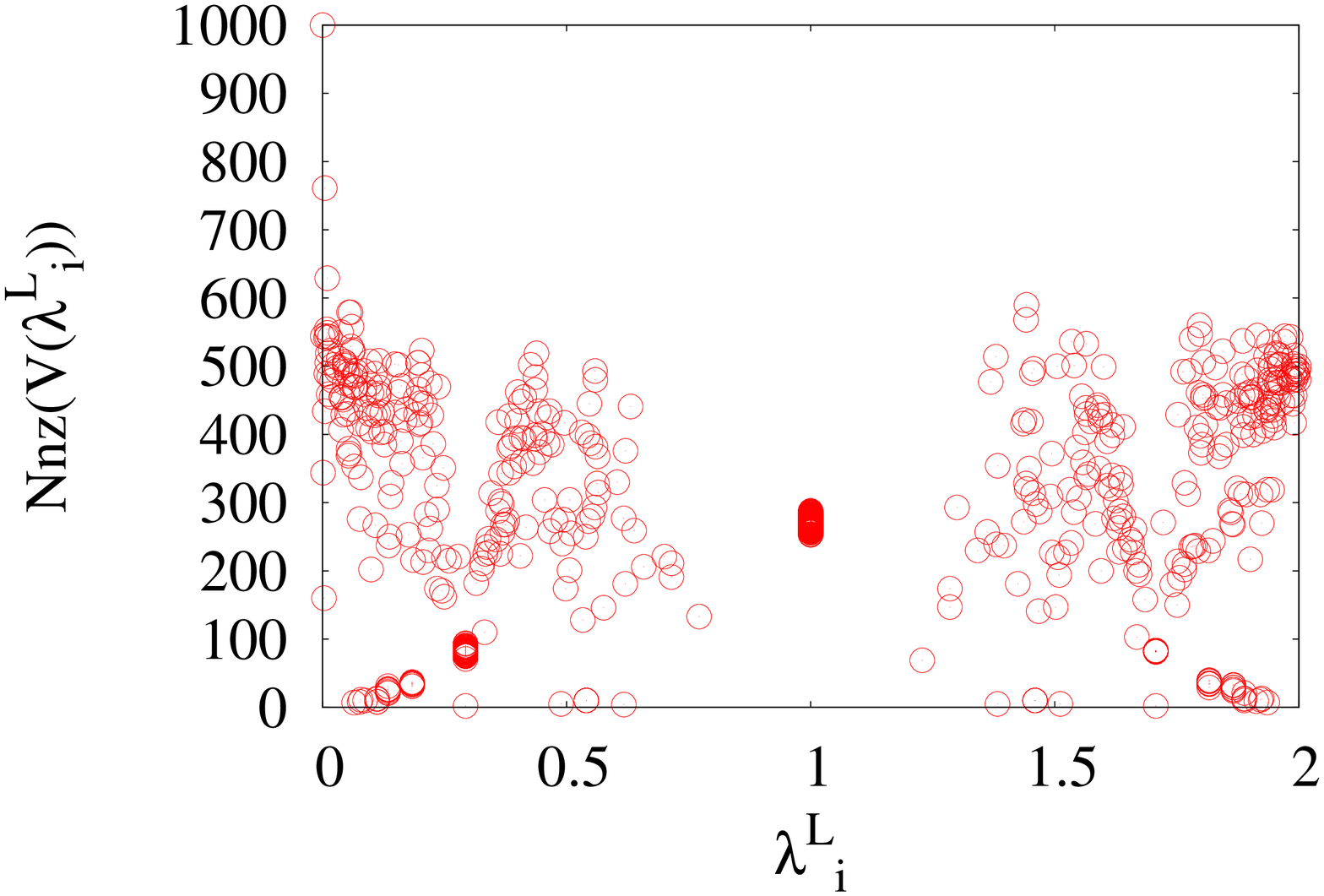}\\
\includegraphics[width=7.2cm]{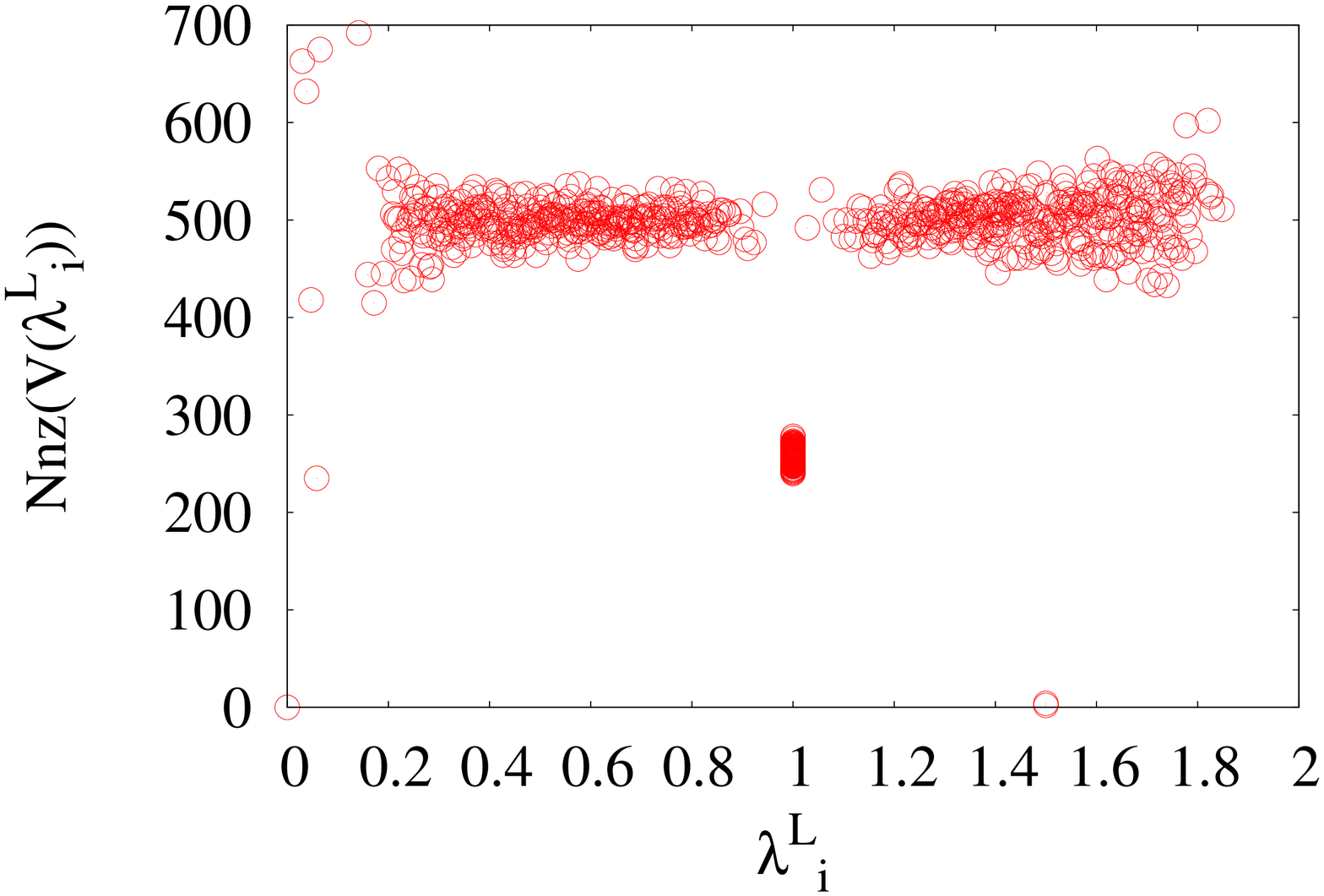}\\
\end{tabular}
\end{center}
\caption{
Number of nodes $Nnz$ carrying a non-zero eigenvector component $V_\kappa(\lambda_i^L)$ plotted against the corresponding eigenvalue $\lambda_i^L$  for the normalized Laplacian of the networks: tree of trees Net161 (top) and cyclic Net269 (bottom).}
\label{participation_ratio}
\end{figure}
A scalar measure of vector's degree of localization is so-called {\it inverse participation ratio} (IPR) \cite{mcgraw2008}, which is defined for any vector $V(\lambda _i^L)$ by the following expression
\begin{equation}
Pr(V(\lambda_{i}^{L}))=\frac{\sum_{\kappa}V_{\kappa}^{4}(\lambda_{i}^{L})}{(\sum_{\kappa}V_{\kappa}^{2}(\lambda_{i}^{L}))^{2}} \ . \label{ipr}
\end{equation}
Depending on the actual situation, IPR ranges from the minimum value $Pr=\frac{1}{N}$, corresponding to the  eigenvector equally distributed on all nodes in the network, to the maximum value  $Pr=1$, in the case when  the eigenvector has only one component different from zero.  Generally, higher values of $Pr(V(\lambda_{i}^{L}))$ are expected for  better localized eigenvectors in subsets of nodes on the network. Note that the actual values of the eigenvector components may vary a lot throughout the network (see Fig.\ \ref{vec_lap} (top)).
Thus it is interesting to consider the number of nodes carrying a non-zero (within the numerical precision) components of the eigenvector ($Nnz$). In Fig.\  \ref{participation_ratio} we show the number of nodes with non-zero eigenvector components for all eigenvectors of 
the  normalized Laplacian in modular network Net269 and tree of trees network, Net161,  plotted against the corresponding eigenvalues $\lambda_i^{L}$.

For the tree  graph most of the vectors have  non-zero components along up to  $50\%$ of nodes, however, they are not equally distributed over nodes. In contrast, the eigenvectors correspond to $\lambda^{L}=1.707107$ and symmetrically to $\lambda^{L}=0.292893$ are homogeneously distributed on approximately $100$ nodes.
 For the eigenvalues close to unity (plateau in Fig.\ \ref{ranking_tree_of_trees}), the IPR varies between  $0.005$ and $0.27$, but the number of  nodes with non-zero vector component remains close to $300$. 
In contrast, in the case of cyclic network Net269, most of the vectors are localized on $50\%$ of the nodes (Fig.\ \ref{participation_ratio} bottom). The exceptions are the lowest eigenvalues, discussed above,  and $\lambda^L=1$, whose eigenvectors are located at 1/4 of the network.

\section{Random walks on trees and cyclic modular networks \label{sec:RW}}
 The observed differences in the spectra of our modular networks are also manifested in random-walk dynamics on them.  
Many stochastic processes in different fields of science have been formulated and studied in terms of random walks \cite{RW-book2004}.  Random walks on networks strictly adhere to the network structure and thus can be used to explore the network topology at different levels \cite{tadic2003,newman-rwcentrality,zhou2003,wwwrw}.
Two fundamental features of the random-walk processes, on which we will focus in the context of the present work, are described by (a) the number of hits of a random walker to network nodes, related to network's transitivity and recurrence; and (b) the distribution of first return times, which is closely related to the distribution of all returns and to the Laplacian spectrum of the network.
 \begin{widetext}
\begin{figure*}[htb]
\begin{center}
\begin{tabular}{cc} 
\large{(a)}& \large{(b)}\\
\resizebox{16.8pc}{!}{\includegraphics{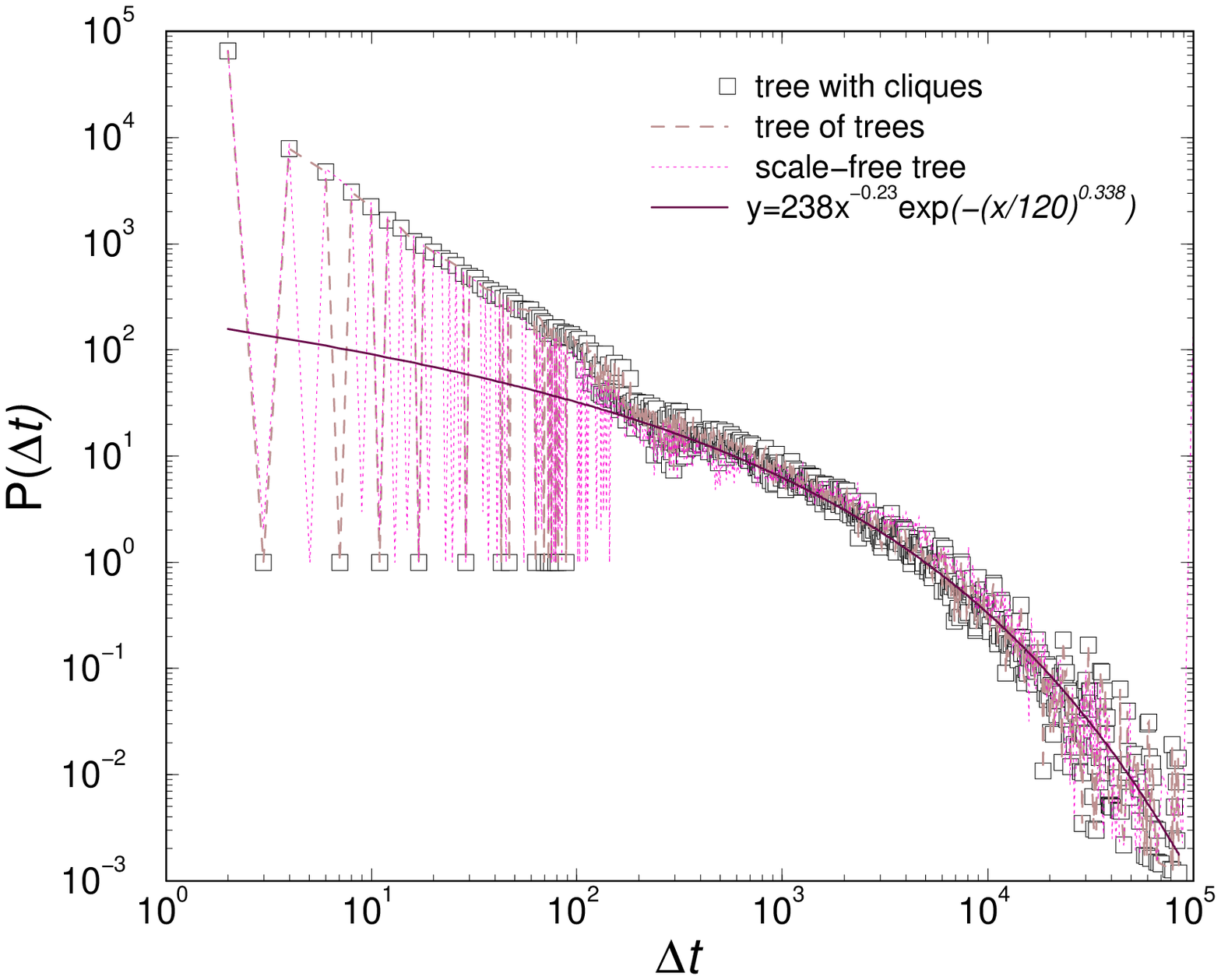}} &
\resizebox{16.8pc}{!}{\includegraphics{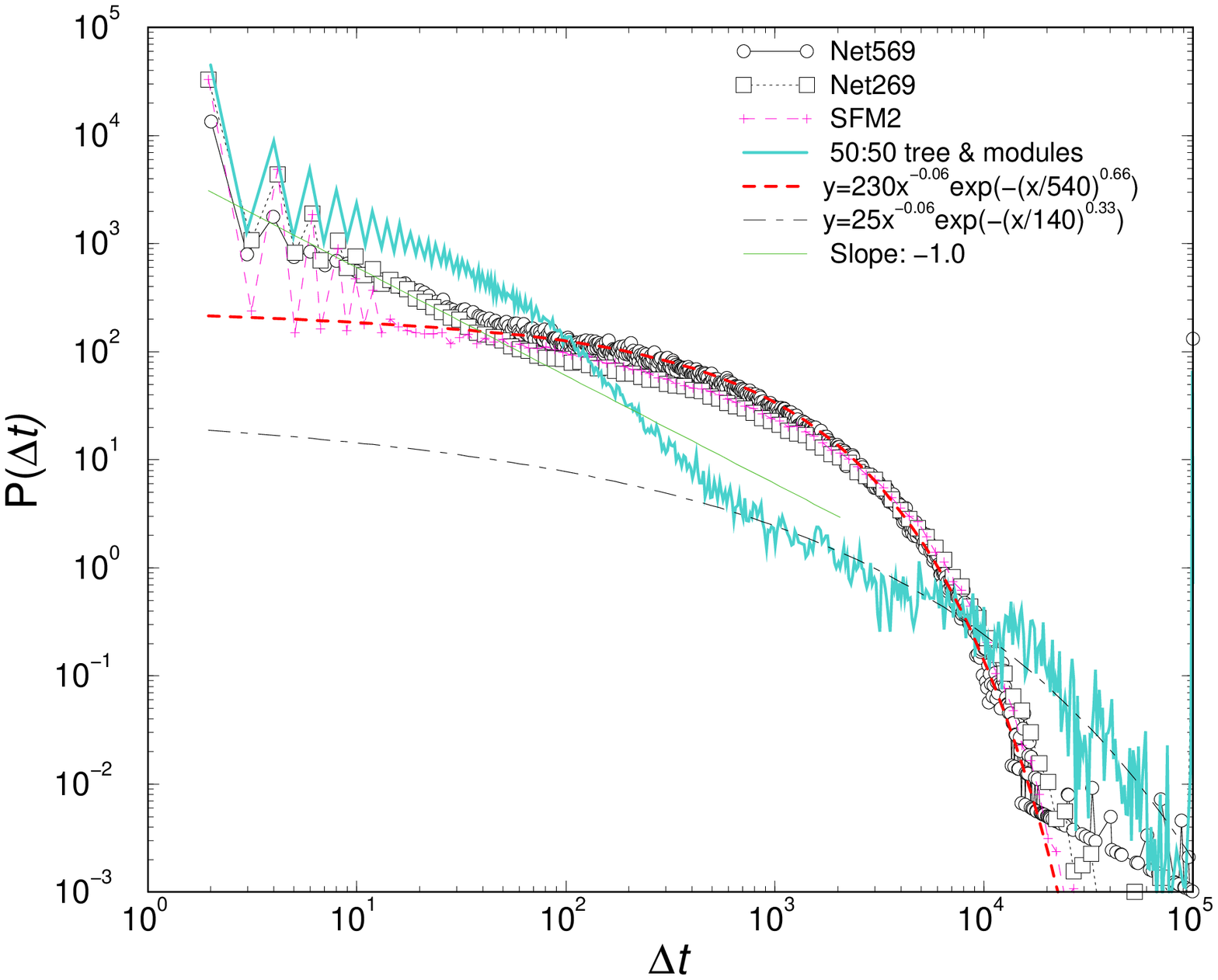}}   \\
\resizebox{16.8pc}{!}{\includegraphics{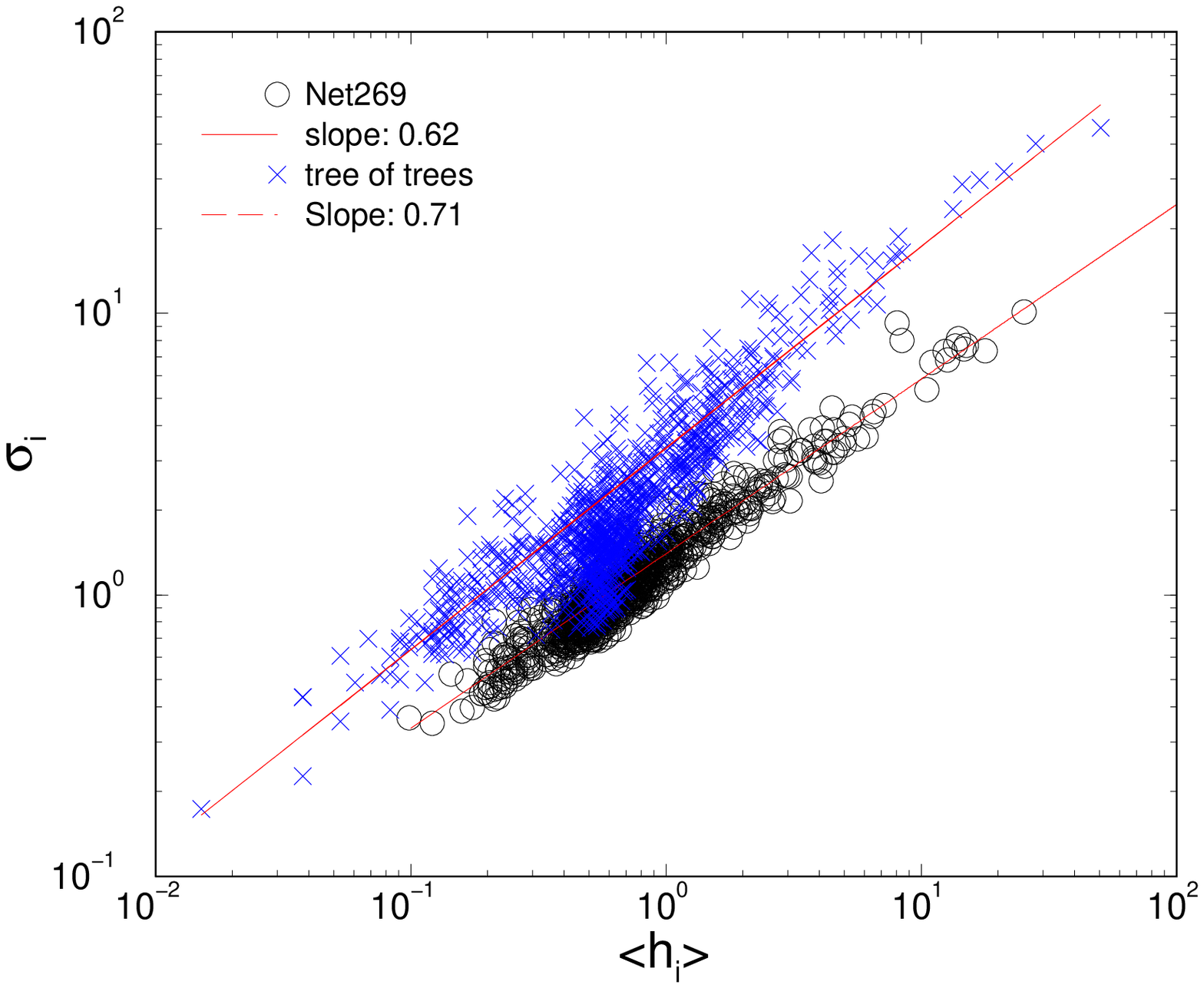}} &
\resizebox{16.8pc}{!}{\includegraphics{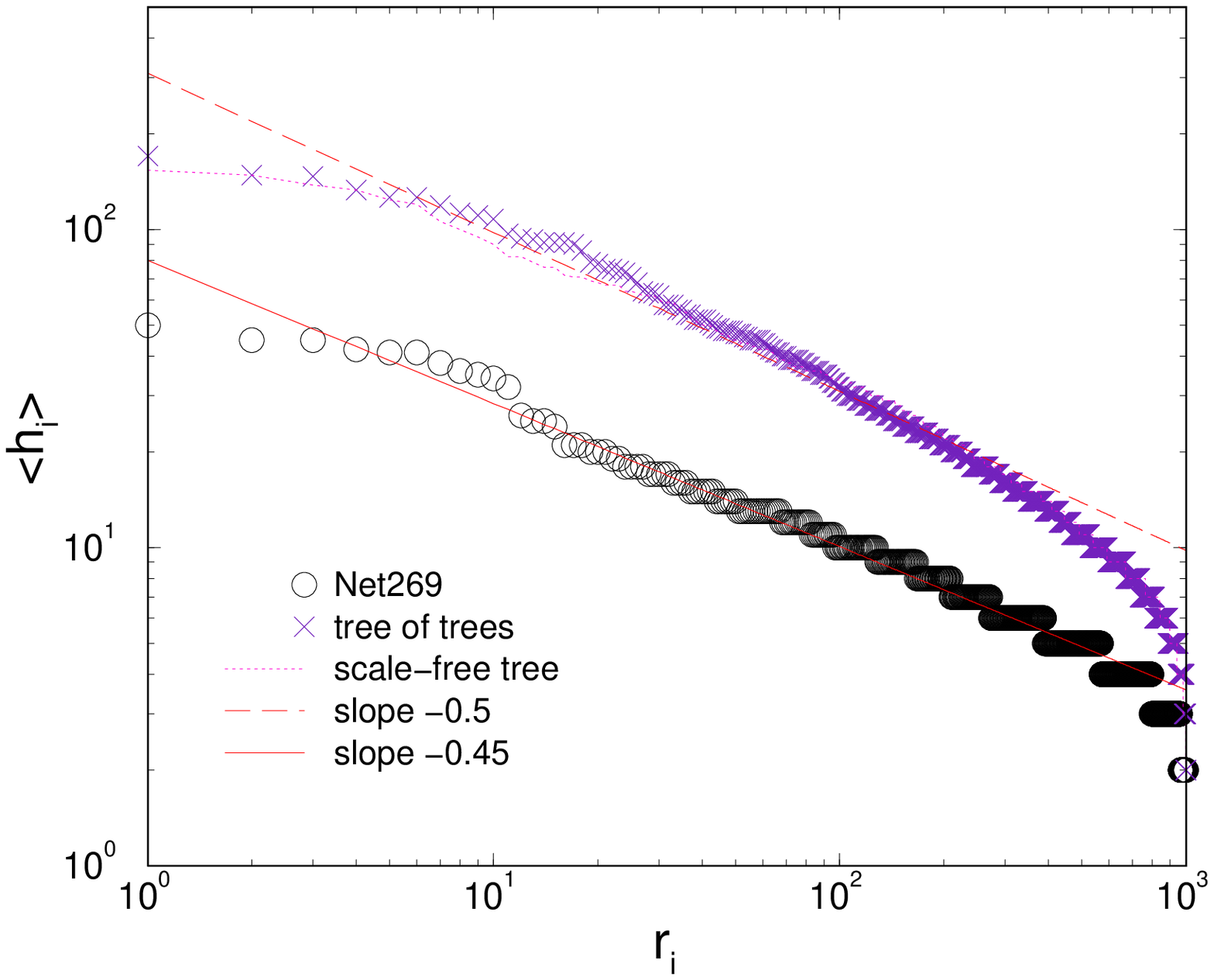}}\\
\large{(c)}&\large{(d)}\\
\end{tabular}
\end{center}
\caption{(Color online.) Statistics of true random walks on (a) trees and tree-like structures and on (b) cyclic modular networks Net269, Net569, and  scale-free network SFM2, all  with  average connectivity $M\geq 2$,  and for mixed network with 50\% of nodes belonging to large modules and 50\% to the connecting tree, shown by thick pale (cyan) line. (c) Dispersion against average of the number of hits time series for all nodes on two networks: tree of trees, and  cyclic network  with  average connectivity M=2 and six modules, Net269. (d) Ranked time-averaged number of hits  per node $\langle h_i\rangle$ versus node rank $r_i$ for two networks as in (c), and for scale-free tree. }
\label{fig-rw1}
\end{figure*} 
\end{widetext}
We simulate the random walks on the networks studied above, Net269 and Net161, and similar structures, which contain $N=1000$ nodes and random walks on each network perform up to $2\times 10^8$ steps. For improved statistics, the total number of walker steps consists of  2000 pieces as follows: First a random walk is started from a node $i$ and when it first arrives to an in advance  randomly selected node $j$, then a new target node $k$ is randomly selected, etc.  In the simulations we measure the elapsed time $\Delta t$ between two successive visits of the walker to the same node (first return to the origin). The times $\Delta t$ are measured at each node of the network and the distribution   $P(\Delta t)$ {\it averaged over all $10^3$ nodes (origins)} and over the ensemble of  2000 walks.

The  probability density functions of  the return times, $P(\Delta t)$ is shown in Fig.\ \ref{fig-rw1} (a) and (b) for different trees and cyclic networks. In addition, we sampled the time series of the number of hits of the random walker  $\{h_i(t)\}$ within a fixed time window of $T_{WIN}=1000$ steps, for each node in the network $i=1,2, \cdots, N$. 
In Fig.\ \ref{fig-rw1}a the results for the first return-times distribution of the walker to a node is shown for the case of of the random walks on scale-free tree, the tree of trees and for the tree with cliques. The numerical results show no significant difference for all kinds of trees (all data are log-binned with a very small base $b=1.01$).
 Moreover,   for $\Delta t >> 1$ the data are well fitted with the expression 
\begin{equation}
P(\Delta t) = B(\Delta t)^{-\eta} \times \exp{\left[-(\Delta t/a)^\sigma \right]} \ 
\label{eq-prett}
\end{equation}
 (shown by full line in Fig.\ \ref{fig-rw1}a), with the exponents $\eta =0.23$ and $\sigma =0.33$ within a numerical error. The presence of small cliques attached to the tree (network shown in Fig.\ \ref{fig-graphs4}d) does not affect the tail of the distribution $P(\Delta t)$ suggesting that long return times are mainly determined by the tree structure of the underlying graph. Note also that there are no significant  difference between the random walks on the random tree and the scale-free tree, as well as the tree of trees (network  in Fig.\ \ref{fig-graphs4}c). In one of the early works considering the diffusion on random  graphs \cite{bray-rodgers1988} the exponent 1/3 for the case of  random walk on a tree was derived by heuristic arguments. 

A similar expression fits the distribution in the case of random walks on cyclic graphs, however, with different exponents. The fit suggests the stretching exponent $\sigma \approx 0.66$, that is twice larger compared with the case of trees. The  simulations of random walks on various cyclic networks, also shown in Fig.\ \ref{fig-rw1}b, suggest that, within the numerical accuracy, the tail of the distribution $P(\Delta t)$ is practically independent on the size of cycles including triangles. 
 A short region with very small slope is found in the intermediate part of the curve, in agreement with Eq.\ (\ref{eq-prett}) with very small $\eta$. For short return times  $\Delta t < 10^2$ we find   tendency to a power-law dependence as $P(\Delta t) \sim (\Delta t)^{-1}$, which is more pronounced in networks 
with increased clustering and modularity. For the tree with attached modules, described in Sec.\ \ref{sec:model} and shown in Fig.\ \ref{fig-graphs4}d, the tail of  return-time distribution $P(\Delta t)$ on this network are also shown in Fig.\ \ref{fig-rw1}b: the tail tends to oscillate between the curves for the trees (long-dashed line) and the cyclic graphs, with a pronounced crossover at short times.

Analysis of the time series $\{h_i(t)\}$ of the number of hits of the random walker to 
 each node reveals additional regularity in the dynamics, which underly the return-time distribution. In Fig.\ \ref{fig-rw1}c we show the scatter plot of the dispersion  $\sigma _i$ of the time series $h_i(t)$  against the average $<h_i(t)>_t$ for each node $i=1,2, \cdots, N$ represented by a point. As shown in Fig.\ \ref{fig-rw1}c, long-range correlations in the diffusion processes on networks lead to a non-universal scaling relation \cite{ktr-njp07}
\begin{equation}
\sigma _i = const \times \langle h_i\rangle_t^\mu \ ,
\label{sigma-h}
\end{equation}
where the averaging over all time windows is taken. The exponent $\mu$ depends on the network structure and the size of the time window. The origin of scaling in diffusive processes on networks has been discussed in detail in \cite{ktr-njp07} and references therein. Here we stress the differences of the underlying networks for the fixed time window $T_{WIN}=1000$: we find $\mu = 0.7$ and $\mu = 0.62$ for tree of trees and modular network Net269, respectively. In this plots the groups of nodes that are most often visited can be identified  at the top-right region of the plot. 

\begin{figure}[htb]
\begin{center}
\begin{tabular}{c} 
\resizebox{20pc}{!}{\includegraphics{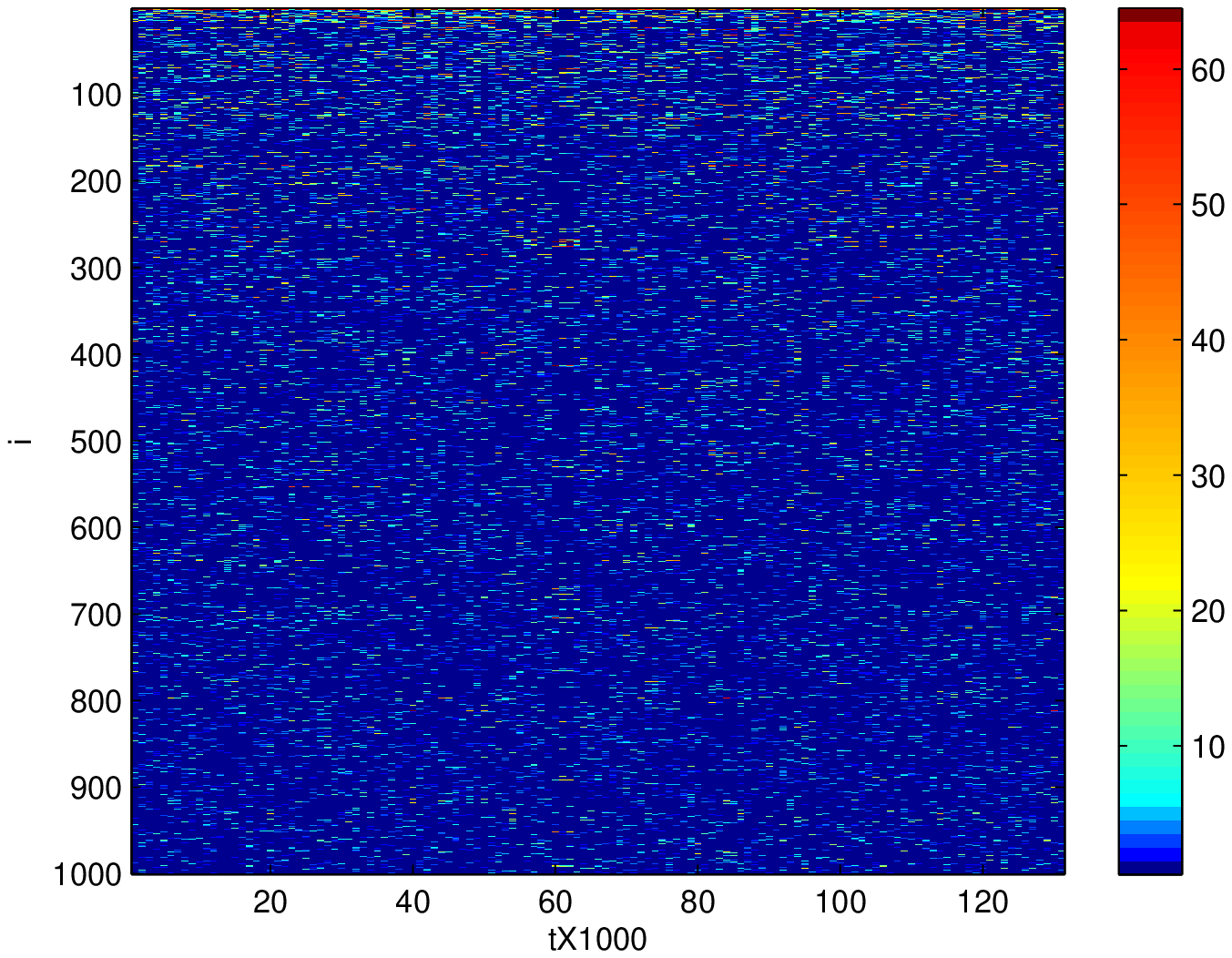}} \\
\resizebox{20pc}{!}{\includegraphics{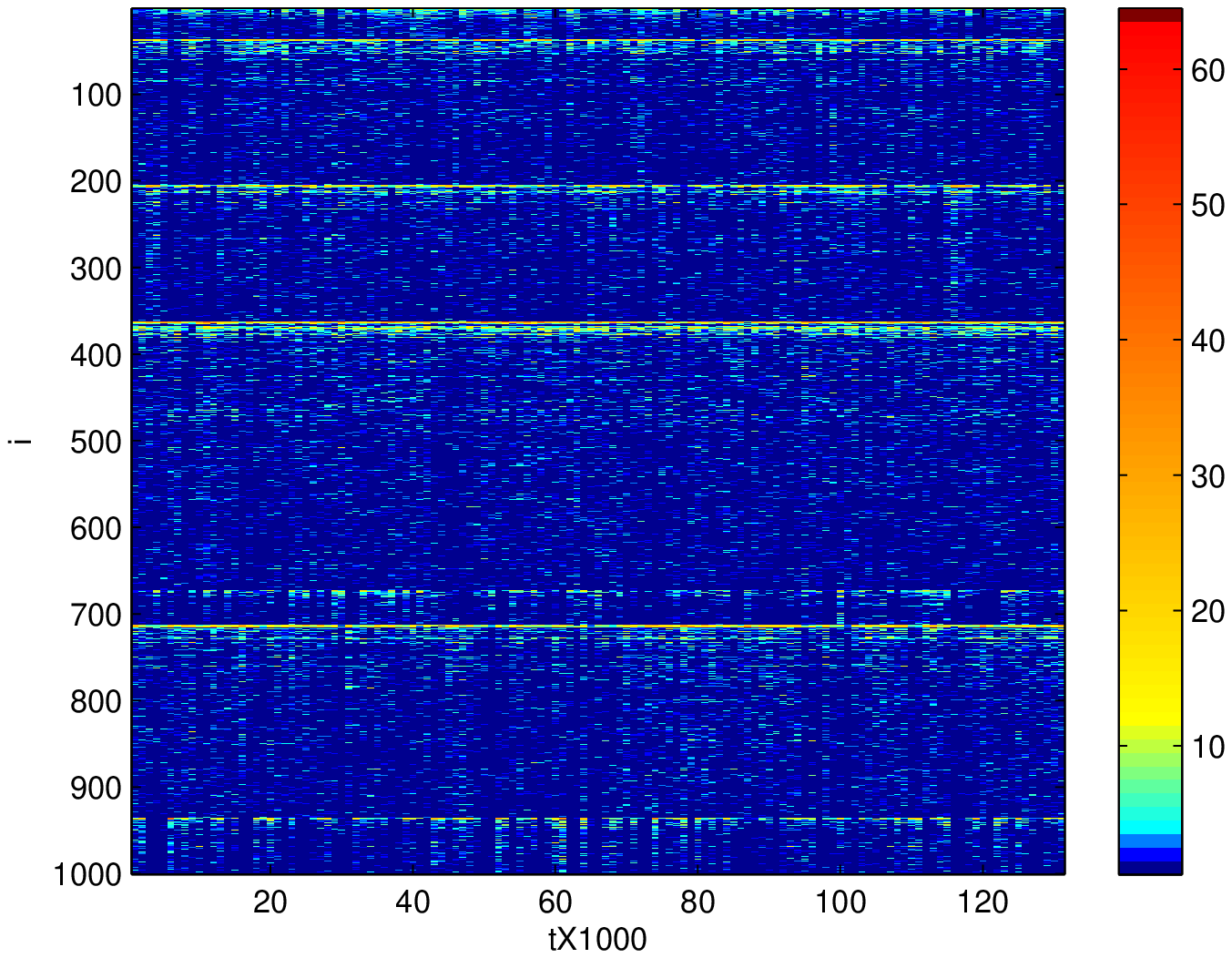}} \\
\end{tabular}
\end{center}
\caption{(Color online.) 3-dimensional plot of the temporal evolution of random walks on  tree of trees, Net161 (top), and on cyclic modular network, Net269 (bottom). Indexes of nodes are shown along the vertical axis, while the horizontal axis indicates 130 time windows, each consisting of 1000 time steps. Color code represents the number of hits of the random walk at the node within the corresponding time window. Brighter/yellow-red color indicates larger number of hits.}
\label{fig-3d}
\end{figure}

Ranking distribution of the average number of hits $<h_i(t)>_t$ at nodes is shown in Fig.\ \ref{fig-rw1}d for the same two representative networks. Generally, the number of hits of true random walker to a node is proportional to node connectivity \cite{tadic2003} and thus, the ranking distribution is a power-law with the slope $\gamma$ which is directly related to the ranking distribution of degree in Fig.\ \ref{degree-ranking}.  
In the presence of network modularity we realize the flat part of the curve, representing most visited nodes. In our modular network, like Net269, these nodes are roots of different modules. A similar feature can be seen in the case of tree of trees  (top curve in Fig.\ \ref{fig-rw1}d). However, closer inspection of the number of hits in time, shown in 3-dimensional plot in Fig.\ \ref{fig-3d}, suggests that the most visited nodes on the trees are not necessarily related to the roots of the subtrees. Whereas, in the cyclic modular Net269 the root nodes of each module can be clearly identified as most visited nodes and are identical to the most connected nodes in each module. 

\section{Conclusions\label{sec:conclusions}}
 We have presented a model of growing modular networks in which we tune the structural properties at all scales by varying the respective control parameters. Specifically, the parameter $P_0$ controls the number of topologically distinct subgraphs (modules); the parameter $\alpha$ is directly related to the number of rewired links which, in turn, determine the clustering inside the scale-free subgraphs and connections between different modules; the parameter $M$ is the average  number of links per node, which can be varied  independently on the clustering and modularity. The wide class of mesoscopically inhomogeneous networks grown by varying these parameters  includes the {\it sparse modular graphs} with variety of topological features, both within the modules and at the level of the connecting network. Two limiting cases are the interacting scale-free networks with finite clustering and correlations, at one end, and 
a scale-free tree supporting a large number of cyclic modules, on the other (see Fig.\ \ref{fig-graphs4}). 

We further study the spectral properties of these networks by focusing on the normalized Laplacian matrix which is  related to the diffusion (random walk) processes on these networks. We have also explored these networks by simulating random walks on them. 
The systematic analysis of the spectra while the network grows and by varying the control parameters enabled us to point out the role played by a specific topological property of the network (controlled by different parameter) in their spectra and the dynamics. Two prototype modular networks---tree connecting scale-free tree subgraphs (Net161) and the cyclic graph connecting scale-free clustered subgraphs (Net269), exhibit systematic differences at the level of spectra and the random-walk dynamics. 

Our complete spectral analysis reveals the firm connection between the structure and spectra. We have found several  new results and also demonstrated  clearly how some expected results for this type of graphs are related to the structure. Specifically, we point out:
\begin{itemize}
 \item{} the role of most connected nodes as opposed to the role of the underlying tree graph;
 
\item{} the role that least connected nodes play in the appearance of the central peak;

\item{} how the increased clustering  changes the shape of the spectrum near the central peak;

\item{} the appearance of the extra peak at small eigenvalues of the normalized Laplacian; In view of our fine tunning of the structure, this peak is related to the number of distinct modules even if the graphs are very sparse, as long as they are cyclic; Increased clustering does not affect this peak;

\item{}the Laplacian spectra of trees are different, particularly, they do not show extra peak related with the (tree) subgraphs; 

\item{} the eigenvector components  show  the expected pattern of localization on modules in the case of small nonzero eigenvalues of the Laplacian. In spite of the differences in the spectral density, we find a similar localization pattern in trees with tree subgraphs.
In addition, we find a robust localization along the network chains of the eigenvectors for largest eigenvalue of the Laplacian $\lambda^L =2$, occurring only in trees and bipartite graphs;

\item{}the simulation results for the first return-time distribution  of the random walks, averaged over network nodes, on different tree graphs exhibits a power-law with stretching exponential cut-off  with $\sigma \approx 1/3$ in agreement with Eq.\ (\ref{eq-prett}) and heuristic arguments \cite{bray-rodgers1988};

\item{} when the graph contains cycles, the distribution  belongs to another class of behavior with twice larger stretching  exponent.  The numerical results  do not dependent on the clustering coefficient.
The presence of modules and increased clustering affect the behavior at small times, where a power-law decay occurs before the cutoff.
\end{itemize}

Our systematic numerical study along the line structure--spectra--random-walks quantifies the relationships between different structural elements of the network and their spectra and the dynamics. 
We hope that the presented  results contribute to better understanding of the diffusion processes on the sparse graphs with complex topology. Potentially, some of our findings may be used for fine  differentiation between  classes of graphs with respect their spectral and dynamical properties.

\acknowledgments
Research supported in part by the program  P1-0044
(Slovenia) and national project OI141035 (Serbia), bilateral project BI-RS/08-09-047 and MRTN-CT-2004-005728.
The numerical results were obtained on the AEGIS
e-Infrastructure, supported in part by EU FP6 and FP7 projects EGEE-III, SEE-GRID-SCI and CX-CMCS.

%\bibliographystyle{revtex}
%\bibliography{biblio_mitrovic_ee.bib}

\begin{thebibliography}{10}
\providecommand*{\bibinfo}[2]{#2}
\providecommand*{\eprint}[1]{#1}
\providecommand*{\url}[1]{#1}
\bibitem{boccaletti2006}
\bibinfo{author}{S.~{Boccaletti}}, \bibinfo{author}{V.~{Latora}},
  \bibinfo{author}{Y.~{Moreno}}, \bibinfo{author}{M.~{Chavez}}, and
  \bibinfo{author}{D.-U. {Hwang}}, \bibinfo{journal}{Physics Reports}
  \bibinfo{volume}{\textbf{424}}, \bibinfo{pages}{175} (\bibinfo{date}{2006}).
\bibitem{tadic2007}
\bibinfo{author}{B.~{Tadi\'c}}, \bibinfo{author}{G.~J. {Rodgers}}, and
  \bibinfo{author}{S.~{Thurner}}, \bibinfo{journal}{International Journal of
  Bifurcation and Chaos} \bibinfo{volume}{\textbf{17}}(7),
  \bibinfo{pages}{2363} (\bibinfo{date}{2007}).
\bibitem{communities}
\bibinfo{author}{A.~{Arenas}}, \bibinfo{author}{L.~{Danon}},
  \bibinfo{author}{A.~{D{\'{\i}}az-Guilera}}, \bibinfo{author}{P.~M.
  {Gleiser}}, and \bibinfo{author}{R.~{Guimer{\`a}}},
  \bibinfo{journal}{European Physical Journal B} \bibinfo{volume}{\textbf{38}},
  \bibinfo{pages}{373} (\bibinfo{date}{2004}).
\bibitem{metabolic}
\bibinfo{author}{E.~Ravasz}, \bibinfo{author}{A.~L. Somera},
  \bibinfo{author}{D.~A. Mongru}, \bibinfo{author}{Z.~N. Oltvai}, and
  \bibinfo{author}{A.~L. Barabasi}, \bibinfo{journal}{Science}
  \bibinfo{volume}{\textbf{297}}, \bibinfo{pages}{1551} (\bibinfo{date}{2002}).
\bibitem{motifs}
\bibinfo{author}{R.~{Milo}}, \bibinfo{author}{S.~{Shen-Orr}},
  \bibinfo{author}{S.~{Itzkovitz}}, \bibinfo{author}{N.~{Kashtan}},
  \bibinfo{author}{D.~{Chklovskii}}, and \bibinfo{author}{U.~{Alon}},
  \bibinfo{journal}{Science} \bibinfo{volume}{\textbf{298}},
  \bibinfo{pages}{824} (\bibinfo{date}{2002}).
\bibitem{neural_nets}
\bibinfo{author}{R.~{Graben}}, \bibinfo{author}{C.~{Zhou}},
  \bibinfo{author}{M.~{Thiel}}, and \bibinfo{author}{J.~{Kurths}},
  \bibinfo{title}{\emph{{Lectures in Supercomputational Neuroscience: Dynamics
  in Complex Brain Networks (Understanding Complex Systems}}}
  (\bibinfo{publisher}{Springer-Verlafg, Berlin Heidelberg},
  \bibinfo{year}{2008}).
\bibitem{Costa2008}
\bibinfo{author}{P.~R. {Villas Boas}}, \bibinfo{author}{F.~A. {Rodrigues}},
  \bibinfo{author}{G.~{Travieso}}, and \bibinfo{author}{L.~{da Fontoura
  Costa}}, \bibinfo{journal}{\pre} \bibinfo{volume}{\textbf{77}}(2),
  \bibinfo{pages}{026106} (\bibinfo{date}{2008}).
\bibitem{multinetworks}
\bibinfo{author}{A.~{Aleksiejuk}}, \bibinfo{author}{J.~A. {Holyst}}, and
  \bibinfo{author}{D.~{Stauffer}}, \bibinfo{journal}{Physica A Statistical
  Mechanics and its Applications} \bibinfo{volume}{\textbf{310}},
  \bibinfo{pages}{260} (\bibinfo{date}{2002}).
\bibitem{ecosystems}
\bibinfo{author}{J.~M. {Olesen}}, \bibinfo{author}{J.~{Bascompte}},
  \bibinfo{author}{Y.~L. {Dupont}}, and \bibinfo{author}{P.~{Jordano}},
  \bibinfo{journal}{Proceedings of the National Academy of Science}
  \bibinfo{volume}{\textbf{104}}, \bibinfo{pages}{19891}
  (\bibinfo{date}{2007}).
\bibitem{danon2006}
\bibinfo{author}{L.~{Danon}}, \bibinfo{author}{A.~{D\'{i}az-Guilera}}, and
  \bibinfo{author}{A.~{Arenas}}, \bibinfo{journal}{Journal of Statistical
  Mechanics: Theory and Experiment} \bibinfo{volume}{\textbf{11}},
  \bibinfo{pages}{P11010} (\bibinfo{date}{2006}).
\bibitem{Vito-dyn-centrality}
\bibinfo{author}{S.~{Fortunato}}, \bibinfo{author}{V.~{Latora}}, and
  \bibinfo{author}{M.~{Marchiori}}, \bibinfo{journal}{\pre}
  \bibinfo{volume}{\textbf{70}}(5), \bibinfo{pages}{056104}
  (\bibinfo{date}{2004}).
\bibitem{newman2007}
\bibinfo{author}{M.~E.~J. {Newman}} and \bibinfo{author}{E.~A. {Leicht}},
  \bibinfo{journal}{Proceedings of the National Academy of Sciences}
  \bibinfo{volume}{\textbf{104}}(23), \bibinfo{pages}{9564}
  (\bibinfo{date}{2007}).
\bibitem{mmbtLNCS}
\bibinfo{author}{M.~{Mitrovi\'c}} and \bibinfo{author}{B.~{Tadi\'c}},
  \bibinfo{journal}{Lecture Notes in Computer Science}
  \bibinfo{volume}{\textbf{5102}}, \bibinfo{pages}{551} (\bibinfo{date}{2008}).
\bibitem{arenas2006}
\bibinfo{author}{A.~{Arenas}}, \bibinfo{author}{A.~{D{\'{\i}}az-Guilera}}, and
  \bibinfo{author}{C.~J. {P{\'e}rez-Vicente}}, \bibinfo{journal}{Physical
  Review Letters} \bibinfo{volume}{\textbf{96}}(11), \bibinfo{pages}{114102}
  (\bibinfo{date}{2006}).
\bibitem{donetti2004}
\bibinfo{author}{L.~{Donetti}} and \bibinfo{author}{M.~A. {Mu{\~n}oz}},
  \bibinfo{journal}{Journal of Statistical Mechanics: Theory and Experiment}
  \bibinfo{volume}{\textbf{10}}, \bibinfo{pages}{P10012},  (\bibinfo{date}{2004}).
\bibitem{cormen2001}
\bibinfo{author}{T.~H. {Cormen}}, \bibinfo{author}{C.~E. {Leiserson}},
  \bibinfo{author}{R.~L. {Rivest}}, and \bibinfo{author}{C.~{Stein}},
  \bibinfo{title}{\emph{{Introduction to Algorithms, 2nd edn.}}}
  (\bibinfo{publisher}{MIT Press and McGraw-Hil}, \bibinfo{year}{2001}).
\bibitem{algebraic_topology_book}
\bibinfo{author}{A.~{Hatcher}}, \bibinfo{title}{\emph{{Algebraic Topology}}}
  (\bibinfo{publisher}{Cambridge University Press}, \bibinfo{year}{2002}).
\bibitem{MilanLNCS}
\bibinfo{author}{S.~{Maleti\'c}}, \bibinfo{author}{M.~{Rajkovi\'c}}, and
  \bibinfo{author}{D.~{Vasiljevi\'c}}, \bibinfo{journal}{Lecture Notes in
  Computer Science} \bibinfo{volume}{\textbf{5102}}, \bibinfo{pages}{568}
  (\bibinfo{date}{2008}).
\bibitem{mcgraw2008}
\bibinfo{author}{P.~N. {McGraw}} and \bibinfo{author}{M.~{Menzinger}},
  \bibinfo{journal}{\pre} \bibinfo{volume}{\textbf{77}}(3),
  \bibinfo{pages}{031102} (\bibinfo{date}{2008}).
\bibitem{Jurgen}
\bibinfo{author}{C.~{Zhou}} and \bibinfo{author}{J.~{Kurths}},
  \bibinfo{journal}{Chaos} \bibinfo{volume}{\textbf{16}}(1),
  \bibinfo{pages}{015104} (\bibinfo{date}{2006}).
\bibitem{albertNJP}
\bibinfo{author}{J.~{Almendral}} and \bibinfo{author}{A.~{Diaz-Guilera}},
  \bibinfo{journal}{New Journal of Physics} \bibinfo{volume}{\textbf{9}}(6),
  \bibinfo{pages}{187} (\bibinfo{date}{2007}).
\bibitem{eigen_cen}
\bibinfo{author}{D.~{Bell}}, \bibinfo{author}{J.~{Atkinson}}, and
  \bibinfo{author}{C.~J.W.}, \bibinfo{journal}{Social Networks}
  \bibinfo{volume}{\textbf{21}}(1), \bibinfo{pages}{1} (\bibinfo{date}{1999}).
\bibitem{bt-arw01}
\bibinfo{author}{B.~{Tadi{\'c}}}, \bibinfo{journal}{European Physical Journal
  B} \bibinfo{volume}{\textbf{23}}, \bibinfo{pages}{221}
  (\bibinfo{date}{2001}).
\bibitem{nr-rwcn-04}
\bibinfo{author}{J.~D. {Noh}} and \bibinfo{author}{H.~{Rieger}},
  \bibinfo{journal}{PRL} \bibinfo{volume}{\textbf{92}}(11),
  \bibinfo{pages}{118701} (\bibinfo{date}{2004}).
\bibitem{ktr-njp07}
\bibinfo{author}{B.~{Kujawski}}, \bibinfo{author}{B.~{Tadi\'c}}, and
  \bibinfo{author}{G.~J. {Rodgers}}, \bibinfo{journal}{New Journal of Physics}
  \bibinfo{volume}{\textbf{9}}, \bibinfo{pages}{154} (\bibinfo{date}{2007}).
\bibitem{random-matrices}
\bibinfo{author}{S.~F. Edwards} and \bibinfo{author}{R.~C. Jones},
  \bibinfo{journal}{Journal of Physics A: Mathematical and General}
  \bibinfo{volume}{\textbf{9}}(10), \bibinfo{pages}{1595}
  (\bibinfo{date}{1976}).
\bibitem{farkas2001}
\bibinfo{author}{I.~J. {Farkas}}, \bibinfo{author}{I.~{Der\'enyi}},
  \bibinfo{author}{A.-L. {Barab\'asi}}, and \bibinfo{author}{T.~{Vicsek}},
  \bibinfo{journal}{Phys. Rev. E} \bibinfo{volume}{\textbf{64}}(2),
  \bibinfo{pages}{026704} (\bibinfo{date}{2001}).
\bibitem{dorogovtsev2003}
\bibinfo{author}{S.~N. {Dorogovtsev}}, \bibinfo{author}{A.~V. {Goltsev}},
  \bibinfo{author}{J.~F. {Mendes}}, and \bibinfo{author}{A.~N. {Samukhin}},
  \bibinfo{journal}{\pre} \bibinfo{volume}{\textbf{68}}(4),
  \bibinfo{pages}{046109} (\bibinfo{date}{2003}).
\bibitem{goh2001}
\bibinfo{author}{K.-I. Goh}, \bibinfo{author}{B.~Kahng}, and
  \bibinfo{author}{D.~Kim}, \bibinfo{journal}{Phys. Rev. E}
  \bibinfo{volume}{\textbf{64}}(5), \bibinfo{pages}{051903}
  (\bibinfo{date}{2001}).
\bibitem{samukhin2007}
\bibinfo{author}{A.~N. {Samukhin}}, \bibinfo{author}{S.~N. {Dorogovtsev}}, and
  \bibinfo{author}{J.~F.~F. {Mendes}}, \bibinfo{journal}{\pre}
  \bibinfo{volume}{\textbf{77}}(3), \bibinfo{pages}{036115}
  (\bibinfo{date}{2008}).
\bibitem{jost2008}
\bibinfo{author}{A.~{Banerjee}} and \bibinfo{author}{J.~{Jost}},
  \bibinfo{journal}{Networks and Heterogeneous Media}
  \bibinfo{volume}{\textbf{3}}(2), \bibinfo{pages}{395} (\bibinfo{date}{2008}).
\bibitem{geoff1988}
\bibinfo{author}{G.~J. Rodgers}, \bibinfo{author}{K.~Austin},
  \bibinfo{author}{B.~Kahng}, and \bibinfo{author}{D.~Kim},
  \bibinfo{journal}{Journal of Physics A: Mathematical and General}
  \bibinfo{volume}{\textbf{38}}(43), \bibinfo{pages}{9431}
  (\bibinfo{date}{2005}).
\bibitem{tadic2001}
\bibinfo{author}{B.~{Tadi\'c}}, \bibinfo{journal}{Physica A Statistical
  Mechanics and its Applications} \bibinfo{volume}{\textbf{293}},
  \bibinfo{pages}{273} (\bibinfo{date}{2001}).
\bibitem{dorogovtsev2000}
\bibinfo{author}{S.~N. Dorogovtsev}, \bibinfo{author}{J.~F.~F. Mendes}, and
  \bibinfo{author}{A.~N. Samukhin}, \bibinfo{journal}{Phys. Rev. Lett.}
  \bibinfo{volume}{\textbf{85}}(21), \bibinfo{pages}{4633}
  (\bibinfo{date}{2000}).
\bibitem{nrc}
\bibinfo{author}{W.~{Press}}, \bibinfo{author}{S.~{Teukolsky}},
  \bibinfo{author}{W.~{Vetterling}}, and \bibinfo{author}{B.~{Flannery}},
  \bibinfo{title}{\emph{{Numerical Recipes in C-The Art of Scientific
  Computing, 2nd edn.}}} (\bibinfo{publisher}{Cambridge University Press},
  \bibinfo{year}{1992}).
\bibitem{newman2003}
\bibinfo{author}{M.~E.~J. Newman}, \bibinfo{journal}{Phys. Rev. E}
  \bibinfo{volume}{\textbf{70}}(5), \bibinfo{pages}{056131}
  (\bibinfo{date}{2004}).
\bibitem{supplementaryPRE08}
\bibinfo{author}{M.~Mitrovi\'c},
  \bibinfo{journal}{http://www.scl.rs/papers/supplemetray.pdf}
  (\bibinfo{date}{2008}).
\bibitem{guimera2002}
\bibinfo{author}{R.~{Guimer{\`a}}}, \bibinfo{author}{A.~{D{\'{\i}}az-Guilera}},
  \bibinfo{author}{F.~{Vega-Redondo}}, \bibinfo{author}{A.~{Cabrales}}, and
  \bibinfo{author}{A.~{Arenas}}, \bibinfo{journal}{Physical Review Letters}
  \bibinfo{volume}{\textbf{89}}(24), \bibinfo{pages}{248701}
  (\bibinfo{date}{2002}).
\bibitem{tadic2005}
\bibinfo{author}{B.~{Tadi\'c}} and \bibinfo{author}{S.~{Thurner}},
  \bibinfo{journal}{Physica A} \bibinfo{volume}{\textbf{346}},
  \bibinfo{pages}{183} (\bibinfo{date}{2005}).
\bibitem{agata2007}
\bibinfo{author}{A.~{Fronczak}} and \bibinfo{author}{P.~{Fronczak}},
  \bibinfo{journal}{arXiv:0709.2231}  (\bibinfo{date}{2007}).
\bibitem{adg2008}
\bibinfo{author}{A.~{Diaz-Guilera}}, \bibinfo{journal}{J. Phys. A: Math.
  Theor.} \bibinfo{volume}{\textbf{41}}, \bibinfo{pages}{224007}
  (\bibinfo{date}{2008}).
\bibitem{RW-book2004}
\bibinfo{author}{V.~A. {Kaymanovich}}, \bibinfo{title}{\emph{{Random Walks and
  Geometry: Proceedings of a Workshop at the Erwin Schr�dinger Institute,
  Vienna}}} (\bibinfo{publisher}{Walter de Gruyter}, \bibinfo{year}{2004}).
\bibitem{tadic2003}
\bibinfo{author}{B.~{Tadi{\'c}}}, in \emph{AIP Conf. Proc. 661: Modeling of
  Complex Systems} (\bibinfo{date}{2003}).
\bibitem{newman-rwcentrality}
\bibinfo{author}{M.~{Newman}}, \bibinfo{journal}{Social networks}
  \bibinfo{volume}{\textbf{27}}, \bibinfo{pages}{39} (\bibinfo{date}{2005}).
\bibitem{zhou2003}
\bibinfo{author}{H.~Zhou}, \bibinfo{journal}{Phys. Rev. E}
  \bibinfo{volume}{\textbf{67}}(6), \bibinfo{pages}{061901} (\bibinfo{date}{Jun
  2003}).
\bibitem{wwwrw}
\bibinfo{author}{J.~{Huang}}, \bibinfo{author}{T.~{Zhu}}, and
  \bibinfo{author}{D.~{Schuurmans}}, \bibinfo{journal}{Lecture Notes in
  Computer Science} \bibinfo{volume}{\textbf{4213}}, \bibinfo{pages}{187}
  (\bibinfo{date}{2006}).
\bibitem{bray-rodgers1988}
\bibinfo{author}{A.~{Bray}} and \bibinfo{author}{G.~{Rodgers}},
  \bibinfo{journal}{Physical Review B} \bibinfo{volume}{\textbf{38}},
  \bibinfo{pages}{11461} (\bibinfo{date}{1988}).

\end{thebibliography}

\end{document}